\documentclass[letter,11pt]{article}
\usepackage{lineno}
\usepackage{amssymb}

\usepackage[margin = 2.50cm]{geometry}
\usepackage{microtype}%if unwanted, comment out or use option "draft"
\usepackage{epsfig}
\usepackage{microtype}
\usepackage{wrapfig}
\usepackage{graphicx}
\usepackage{enumitem}
\usepackage{epsfig,enumerate,amsmath,amsfonts,amssymb,amsthm,mathrsfs,ifpdf}
\usepackage{indentfirst,relsize}
\usepackage[numbers]{natbib}
\usepackage{setspace}
\usepackage{enumerate}
\usepackage{latexsym}
\usepackage{stackrel}
\usepackage[all]{xy}
\usepackage[usenames,dvipsnames]{pstricks}
\usepackage{pst-grad} % For gradients
\usepackage{pst-plot} % For axes
\usepackage{xspace}
\usepackage{bbm}

\usepackage[super]{nth}
\usepackage{hyperref}

\usepackage[T1]{fontenc}
\usepackage{lineno}
\newcommand{\size}[1]{\left| #1 \right|}

\newcommand{\E}{\mathbb{E}}
\newcommand{\remove}[1]{}

\newcommand{\cB}{\mathcal{B}}
\newcommand{\cC}{\mathcal{C}}
\newcommand{\cD}{\mathcal{D}}

\newcommand{\cF}{\mathcal{F}}

\newcommand{\Oh}{\mathcal{O}}

\newcommand{\tOh}{\widetilde{{\mathcal O}}}

\newcommand{\cH}{\mathcal{H}}

\newcommand{\eps}{\varepsilon}
\newcommand{\pr}{\mathbb{P}}

\newcommand{\complain}[1]{\textcolor{red}{#1}}

\newcommand{\spcl}{\rho}

\theoremstyle{plain}
\newtheorem{theo}{Theorem}[section]
\newtheorem{lem}[theo]{Lemma}
\newtheorem{pre}[theo]{Proposition}
\newtheorem{coro}[theo]{Corollary}

\newtheorem{cl}[theo]{Claim}
\theoremstyle{definition}
\newtheorem{defi}[theo]{Definition}
\newtheorem{rem}[theo]{Remark}
\newtheorem{question}{Question}
\newtheorem{obs}[theo]{Observation}

\newcommand{\maxcut}{\mbox{{\sc MaxCut}} }
\newcommand{\tolbip}{$\mbox{{\sc Tol-Bip-Dist}}(G,\eps)$\xspace}

\usepackage{multirow}

\title{Tolerant Bipartiteness Testing in Dense Graphs
}

\author{Arijit Ghosh{\footnote{Indian Statistical Institute, Kolkata, India.}}
\and
Gopinath Mishra{\footnote{University of Warwick, Coventry, UK.}}
\and
Rahul Raychaudhury{\footnote{Duke University, USA.}}
\and
Sayantan Sen{\footnote{Indian Statistical Institute, Kolkata, India.}}
}

\date{}
\begin{document}

\maketitle
\thispagestyle{empty} 

\begin{abstract}
Bipartite testing has been a central problem in the area of property testing since its inception in the seminal work of Goldreich, Goldwasser and Ron [FOCS'96 and JACM'98]. Though the non-tolerant version of bipartite testing has been extensively studied in the literature, the tolerant variant is not well understood. In this paper, we consider the following version of tolerant bipartite testing:
Given a parameter $\varepsilon \in (0,1)$ and access to the adjacency matrix of a graph $G$, we can decide whether $G$ is $\varepsilon$-close to being bipartite or $G$ is at least $(2+\Omega(1))\varepsilon$-far from being bipartite, by performing $\widetilde{\mathcal{O}}\left(\frac{1}{\varepsilon ^3}\right)$ queries and in $2^{\widetilde{\mathcal{O}}(1/\varepsilon)}$ time. This improves upon the state-of-the-art query and time complexities of this problem of $\widetilde{\mathcal{O}}\left(\frac{1}{\varepsilon ^6}\right)$ and  $2^{\widetilde{\mathcal{O}}(1/\varepsilon^2)}$, respectively, from the work of Alon, Fernandez de la Vega, Kannan and Karpinski (STOC'02 and JCSS'03), where $\widetilde{\mathcal{O}}(\cdot)$ hides a factor polynomial in $\log \frac{1}{\varepsilon}$.\\
%The algorithm and analysis in this paper are involved unlike most of works in property testing, where the algorithms are often simple, but their analysis are tricky.
%\comments{to be edited}


\end{abstract}
%\newpage

\section{Introduction}\label{sec:intro}

The field of property testing refers to the model where the main goal is to design efficient algorithms that can decide even without looking into the input at its entirety. Over the past few years, the field have had a very rapid growth, and several interesting techniques and results have emerged. See, e.g., Goldreich~\cite{goldreich2017introduction} for an introduction to property testing.

The field of graph property testing was first introduced in the seminal work of Goldreich, Goldwasser and Ron~\cite{DBLP:journals/jacm/GoldreichGR98}. In that work, the authors studied various interesting and important problems in dense graphs and testing bipartiteness was one of them. Given a dense graph $G$ as an input, the problem is to decide if $G$ is bipartite, or we need to modify at least $\eps n^2$ many entries of the adjacency matrix of $G$ to make it bipartite, using as few queries to the adjacency matrix of $G$ as possible, where $\eps \in (0,1)$ is a proximity parameter.

%However all the works till now have only considered the non-tolerant variant of the problem. In this work, we study the problem in tolerant setting, where given two proximity parameters $\eps_1$ and $\eps_2$ with $0 \leq \eps_1 < \eps_2 \leq 1$, the goal is to decide whether $d_{bip}(G) \leq \eps_1 n^2$ or $d_{bip}(G) \geq \eps_2 n^2$ by performing as few queries as possible. We give an algorithm that can decide whether $d_{bip}(G) \leq \eps n^2$ or $d_{bip}(G) \geq (2+ \Omega(1))\eps n^2$, by performing $\widetilde{\Oh}(\frac{1}{\eps^3})$ many adjacency queries to $G$, with probability at least $\frac{2}{3}$. 

Due to the fundamental nature of the problem, \emph{bipartite testing} has been extensively studied over the past two decades~\cite{DBLP:journals/jacm/GoldreichGR98}. Though there are several works on non-tolerant testing of various graph properties across all models in graph property testing~\cite{DBLP:journals/jacm/GoldreichGR98, goldreich1999sublinear,czumaj2019planar}, there are very few works related to their tolerant counterparts (See Goldreich~\cite{goldreich2017introduction} for an extensive list of various results). To the best of our knowledge, this is the first time tolerant bipartite testing has been \emph{explicitly} studied in the literature.

%\emph{bipartite testing} has been a central problem in the area~\cite{}.

Now we formally define the notion of {\em bipartite distance} and state our main result. Then we discuss our result vis-a-vis the related work.

\begin{defi}[Bipartite distance]\label{defi:bipdist}
A \emph{bipartition} of (the vertices of) a graph $G$ is a function $f:V(G)\rightarrow \{L,R\}$~\footnote{ $L$ and $R$ denote left and right respectively.}. The \emph{bipartite} distance of $G$ with respect to the bipartition $f$ is denoted and defined as 
$$d_{bip}(G,f) : = \left[\sum_{v\in V:f(v)=L}|N(v)\cap f^{-1}(L)|+ \sum_{v\in V:f(v)=R}|N(v)\cap f^{-1}(R)|\right].$$
\end{defi}

Here $N(v)$ denotes the neighborhood of $v$ in $G$. Informally, $d_{bip}(G,f)$ measures the distance of the graph $G$ from being bipartite, with respect to the bipartition $f$. The \emph{bipartite distance} of $G$ is defined as the minimum bipartite distance of $G$ over all possible bipartitions $f$ of $G$, that is, 
$$d_{bip}(G) : = \min_{f}d_{bip}(G,f).$$

Now we are ready to formally state our result.
%Our result is formally stated in the following theorem:
\begin{theo}[Main result]\label{theo:mainoverview}
Given query access to the adjacency matrix of a dense graph $G$ with $n$ vertices and a proximity parameter $\eps \in (0,1)$, there exists an algorithm that, with probability at least $\frac{9}{10}$, decides whether $d_{bip}(G) \leq \eps n^2$ or $d_{bip}(G) \geq (2+\Omega(1)) \eps n^2$, by sampling $\Oh\left(\frac{1}{\eps^3} \log \frac{1}{\eps}\right)$ many vertices in $2^{\Oh\left(\frac{1}{\eps} \log \frac{1}{\eps}\right)}$ time, and performs $\Oh\left(\frac{1}{\eps^3}\log^2 \frac{1}{\eps}\right)$ many queries.

%given adjacency query access to a graph $G$  such that, ,  to the adjacency matrix of $G$. 
\end{theo}

% We will present the algorithm in Section~\ref{sec:algosec}, and its analysis in Section~\ref{sec:correctness}.

% \textcolor{blue}{Need to add something here.}

\subsection{Our result in the context of literature}

Non-tolerant bipartite testing refers to the problem where we are given query access to the adjacency matrix of an unknown graph $G$ and a proximity parameter $\eps \in (0,1)$, and the objective is to decide whether $d_{bip}(G)=0$ or $d_{bip}(G)\geq \eps n^2$. The problem of non-tolerant bipartite testing in the dense graph model was first studied in the seminal work of Goldreich, Goldwasser and Ron~\cite{DBLP:journals/jacm/GoldreichGR98}, and they showed that it admits an algorithm with query complexity $\tOh\left(\frac{1}{\eps^3}\right)$. Later,
Alon and Krivelevich~\cite{alon2002testing} improved the query complexity of the problem to $\tOh\left(\frac{1}{\eps^2}\right)$. They further studied the problem of testing $c$-colorability of dense graph. Note that bipartite testing is a special case of testing $c$-colorability, when $c=2$. They proved that $c$-colorability can be tested by performing $\tOh\left(\frac{1}{\eps^4}\right)$ many queries, for $c \geq 3$. This bound was later improved to $\tOh\left(\frac{1}{\eps^2}\right)$ by Sohler~\cite{DBLP:conf/focs/Sohler12}. On the other hand, for non-tolerant bipartite testing, Bogdanov and Trevisan~\cite{DBLP:conf/coco/BogdanovT04} proved that $\Omega(\frac{1}{{\eps}^2})$ and $\Omega(\frac{1}{{\eps}^{{3}/{2}}})$ many adjacency queries are required by any non-adaptive and adaptive testers, respectively. Later, Gonen and Ron~\cite{gonen2007benefits} further explored the power of adaptive queries for bipartiteness testing. Bogdanov and Li~\cite{bogdanov2010better} showed that bipartiteness can be tested with one-sided error in $\Oh(\frac{1}{\eps^c})$ queries, for some constant $c < 2$, assuming a conjecture~\footnote{The conjecture is stated as follows: if the graph $G$ is $\eps$-far from bipartite, then the induced subgraph of $\widetilde{\Oh}(\frac{1}{\eps})$ vertices would be $\widetilde{\Omega}(\eps)$-far from being bipartite.}. 

Though the non-tolerant variant of bipartite testing is well understood, the query complexity of tolerant version (even for restricted cases like we consider in Theorem~\ref{theo:mainoverview}) has not yet been addressed in the literature. From the result of Alon, Vega, Kannan and Karpinski~\cite{DBLP:journals/jcss/AlonVKK03},for estimating \maxcut~\footnote{\maxcut of a graph $G$ denotes the size of the largest \emph{cut} in $G$.} for any given $\eps$ ($0 < \eps < 1$), it implies that that the bipartite distance of a (dense) graph $G$ can be estimated upto an additive error of $\eps n^2$, by performing $\tOh\left(\frac{1}{\eps^6}\right)$ many queries (see Appendix~\ref{sec:alon} for details, and in particular, see Corollary~\ref{coro:one}). Even for the tolerant version that we consider in Theorem~\ref{theo:mainoverview}, their algorithm does not give any bound better than $\tOh\left(\frac{1}{\eps^6}\right)$. Note that Alon, Vega, Kannan and Karpinski~\cite{DBLP:journals/jcss/AlonVKK03} improved the result of Goldreich, Goldwasser and Ron~\cite{DBLP:journals/jacm/GoldreichGR98}, who had proved that \maxcut can be estimated with an additive error of $\eps n^2$ by performing $\tOh(\frac{1}{\eps^7})$ queries and in time $2^{\tOh(\frac{1}{\eps^3})}$. Though we improve the bound for tolerant bipartite testing (for the restricted case as stated in Theorem~\ref{theo:mainoverview}) substantially from the work of Alon et al.~\cite{DBLP:journals/jcss/AlonVKK03}, we would like to note that this is the first work that studies tolerant bipartite testing explicitly.% \textcolor{blue}{Besides, most of the algorithms related to bipartite testing in the literature do not use any refined technique~\footnote{The algorithms mostly samples a random induced subgraph with number of vertices above a certain threshold and then test whether the induced subgraph that has been sampled is bipartite or not. In general their analysis are often trickier.}, as it is mostly the case in property testing. However, our algorithm as well as its analysis are quiet involved.}

\subsection{Other related works}
Apart from the dense graph model, this problem has also been studied in other models of property testing.
Goldreich and Ron~\cite{goldreich1999sublinear} studied the problem of bipartiteness testing for bounded degree graphs, where they gave an algorithm of $\widetilde{\Oh}(\sqrt{n})$ queries, where $n$ denotes the number of vertices of the graph. Later, Kaufman, Krivelevich and Ron~\cite{kaufman2004tight} studied the problem in the general graph model and gave an algorithm with query complexity $\tOh(\min(\sqrt{n}, \frac{n^2}{m}))$, where $m$ denotes the number of edges of the graph. Few years back, Czumaj, Monemizadeh, Onak and Sohler~\cite{czumaj2019planar} studied the problem for planar graphs (more generally, for any minor-free graph), where they employed random walk based techniques, and proved that constant number of queries are enough for the same. {Apart from bipartite testing, there have been extensive works related to property testing in the dense graph model and its connection to the regularity lemma~\cite{alon2009combinatorial,alon2000efficient,fischer2007testing}.}

%like bounded degree graph model, general graph model~ and planar graphs settings~\cite{czumaj2019planar}.

%the We describe the bottleneck of our technique to extend Theorem~\ref{} for bipartite distance estimation in Section~\ref{}.

%We will write about non-tolerant testing.

\color{black}

%\end{rem}
\remove{
\begin{rem}
Note that the above theorem is a result on tolerant bipartite testing. In the non-tolerant version, given a parameter $\delta$, we want to decide whether $d_{bip}(G)=0$ or $d_{bip}(G)\geq \delta n^2$. This version can be solved by using Theorem~\ref{theo:mainoverview} with suitable parameter $\eps$ where $\eps=\frac{\delta}{2+\Omega(1)}$.
\end{rem}

{\sc Tolerant-Bipartite-Testing}$(\eps_1,\eps_2)$: In this problem, the algorithm has query access to the adjacency matrix of a unknown graph $G$, takes two parameters $\eps_1,\eps_2 \in (0,1)$ such that $\eps_1 \leq \eps_2$, and decides whether $d_{bip}(G)\leq \eps_1 n^2$ or $d_{bip}(G)\geq \eps_2n^2$.

{\sc Tolerant-Bipartite-Testing}$(\eps_1,\eps_2)$ problem is said to {\sc Non-tolerant-Bipartite-Testing}$(\eps)$ when $\eps_1=0$ and $\eps_2=\eps$.
\begin{rem}
\end{rem}

}

\subsection{Organization} 

In Section~\ref{sec:pfoverview}, we present an overview of our algorithm along with a brief description of its analysis. In Section~\ref{sec:algosec}, we formally describe our algorithm, followed by its correctness analysis in Section~\ref{sec:correctness}. Finally, we conclude in Section~\ref{sec:conclusion}. The proofs that are omitted in the main text are presented in the appendix.

%Finally, we discuss some open problems related to this work in Section~\ref{sec:conclusion}

\subsection{Notations}

All graphs considered here are undirected, unweighted, and have no self-loops or
parallel edges. For a graph $G(V, E)$, $V(G)$ and $E(G)$ denote the vertex set and the edge set
of $G$ respectively. $N_{G}(v)$ denotes the neighborhood of $v$ in $G$, and we will write it as $N(v)$ when the graph $G$ is clear from the context. Since we are only considering undirected graphs, we write an edge as $\{u, v\} \in E(G)$. For a set of pairs of vertices $Z$, we will denote the set of vertices present in at least one pair in $Z$ by  $V(Z)$. For a function $f:V(G) \rightarrow \{L,R\}$, $f^{-1}(L)~$ $(f^{-1}(R))$ represents the set  of vertices that are mapped to $L~(R)$ by $f$.
${V(G) \choose 2}$ denotes the set of unordered pairs of the vertices of $G$.
Finally,
$a=(1 \pm \eps)b$ represents $(1-\eps)b \leq a \leq (1+\eps)b$.

\section{Overview of the proof of Theorem~\ref{theo:mainoverview} }\label{sec:pfoverview}

%\color{blue}

%Before directly proceeding to the overview of our algorithm, let us first formally define  the notion of bipartite distance which will be crucial for our algorithm in Section~\ref{sec:algosec} and its analysis in Section~\ref{sec:correctness}. 

%\textcolor{red}{Arijit: I will edit this section.}

In this section, we give an overview of our algorithm. The detailed description of the algorithm  is presented in Section~\ref{sec:algosec}, while its analysis is presented in Section~\ref{sec:correctness}. We will prove the following theorem, which is our main technical result. 
\begin{theo}\label{theo:mainoverview_algo}
There exists an algorithm \tolbip that given adjacency query access to a dense graph $G$ with $n$ vertices and a parameter $\eps \in \left(0,1\right)$, decides with probability at least $\frac{9}{10}$, whether $d_{bip}(G) \leq \eps n^2$ or $d_{bip}(G) \geq (2+k) \eps n^2$, by sampling $\Oh(\frac{1}{k^5\eps^2} \log \frac{1}{k\eps})$ many vertices in $2^{\Oh(\frac{1}{k^3\eps} \log \frac{1}{k\eps})}$ time, using $\Oh(\frac{1}{k^8\eps^3}\log^2 \frac{1}{k\eps})$ many queries to the adjacency matrix of $G$.
\end{theo}

Note that Theorem~\ref{theo:mainoverview_algo} implies Theorem~\ref{theo:mainoverview}, assuming $k=\Omega(1)$. 

%in Theorem~\ref{theo:mainoverview}.

\color{black}

\subsection{Brief description of the algorithm}

Assume $C_1, C_2, C_3$ are three suitably chosen large absolute constants.
At the beginning of our algorithm, we generate $t$ many subsets of vertices $X_1,\ldots, X_t$, each with $\lceil{\frac{C_2}{k^3\eps} \log \frac{1}{k\eps}}\rceil$ many vertices chosen randomly, where $t= \lceil{\log \frac{C_1}{k\eps}}\rceil$. Let $\cC=X_1 \cup \ldots \cup X_t$. Apart from the $X_i$'s, we also randomly select a set of pairs of vertices $Z$, with $\size{Z} =\lceil{\frac{C_3}{k^5\eps^2} \log \frac{1}{k \eps}}\rceil$. We find the neighbors of each vertex of $Z$ in $\cC$. Then for each vertex pair in $Z$, we check whether it is an edge in the graph or not. Loosely speaking, the set of edges between $\cC$ and $V(Z)$~\footnote{Recall that $V(Z)$ denotes the set of vertices present in at least one pair in $Z$.} will help us generate partial bipartitions, restricted to $X_i \cup V(Z)$'s, for each $i \in [t]$, and the edges among the pairs of vertices of $Z$ will help us in estimating the bipartite distance of some {\em specific kind} of bipartitions of $G$. Here we would like to note that no further query will be performed by the algorithm. The set of edges with one vertex in $\cC$ and the other in $V(Z)$, and the set of edges among the vertex pairs in $Z$, when treated in a \emph{specific} and non-trivial manner, will give us the desired result. Observe that the number of adjacency queries performed by our algorithm is $\Oh(\frac{1}{k^8\eps^3} \log^2 \frac{1}{k\eps})$.

%\remove{~\footnote{The reason for taking multiple $X_i$'s is just to let the probability calculation work.}}

For each $i\in [t]$, we do the following. We consider all possible bipartitions $\cF_i$ of $X_i$. For each bipartition $f_{ij}$ (of $X_i$)  in $\cF_i$, we extend $f_{ij}$ to a bipartition of $X_i \cup V(Z)$, say $f_{ij}'$, such that both $f_{ij}$ and $f_{ij}'$ are identical with respect to $X_i$. Moreover, we assign $f_{ij}'(z)$ (to either $L$ or $R$), for each $z \in V(Z) \setminus X_i$, based on the neighbors of $z$ in $X_i$.  To design a rule of assigning $f_{ij}'(z)$, for each $z \in V(Z)\setminus X_i$ for our purpose, we define the notions of \emph{heavy} and \emph{balanced} vertices, with respect to a bipartition (see Definition~\ref{defi:heavyvertex} and Definition~\ref{defi:balancedvertex}). Heavy and balanced vertices are defined in such a manner that when the bipartite distance of $G$ is at most $\eps n^2$ (that is, $G$ is $\eps$-close), we can infer the following interesting connections. Let $f$ be a bipartition of $V(G)$ such that $d_{bip}(G,f) \leq \eps n^2$. We will prove that the total number of edges, with no endpoints in $X_i$ and whose at least one end point is a balanced vertex with respect to $f$, is bounded (see Claim~\ref{cl:boundbalanced}). Moreover, if we generate a bipartition $f'$ such that $f$ and $f'$ differ for \emph{large} number of heavy vertices, then the bipartite distance with respect to $f'$ cannot be bounded. To guarantee the correctness of our algorithm, we will prove that a heavy vertex $v$ with respect to $f$, can be detected and $f(v)$ can be determined, with probability at least $1-o(k \eps)$. Note that the testing of being a heavy vertex will be performed only for the vertices in $V(Z)$. We will see shortly how this will help us to guarantee the completeness of our algorithm.

%We will prove that the total number of edges, with no endpoints in $X_i$ and whose at least one end point is a balanced vertex with respect to $f$, is bounded by $12 d_{bip}(G,f) + \frac{\eps n^2}{10}$ (Claim~\ref{cl:boundbalanced}).

Finally, our algorithm computes $\zeta_{ij}$, that is, the fraction of vertex pairs in $Z$ that are monochromatic~\footnote{An edge is said to be monochromatic with respect to $f_{ij}'$ if both its endpoints have the same $f_{ij}'$ values.} edges with respect to $f_{ij}'$. If we find at least one $i$ and $j$ such that $\zeta_{ij} \leq \left(2+\frac{k}{20}\right)\eps$, the algorithm decides that $d_{bip}(G)\leq \eps n^2$. Otherwise, it will report that $d_{bip}(G) \geq (2+ k)\eps n^2$.

%Now we give an overview of the completeness and soundness proofs of our algorithm.  Section~\ref{sec:correctness1}.

\subsection{Completeness}

Let us assume that the bipartite distance of $G$ is at most $\eps n^2$, and let $f$ be a bipartition of $V(G)$ that is optimal. Let us now focus on a particular $i \in [t]$, that is, an $X_i$. Since we are considering all possible bipartitions $\cF_i$ of $X_i$, there exists a $f_{ij} \in \cF_i$, such that $f_{ij}$ and $f$ are identical with respect to $X_i$. To complete our argument, we introduce (in Definition~\ref{defi:splbijection}) the notion of \emph{{\sc special}} bipartition $\mbox{{\sc Spl}}_i^f:V(G) \rightarrow \{L,R\}$, with respect to $f$ by $f_{ij}$ such that $f(v)$, $f_{ij}(v)$ and $\mbox{{\sc Spl}}_i^f(v)$ are identical for each $v \in X_i$, and  at least $1-o(k\eps)$ fraction of heavy vertices, with respect to $f$, are mapped identically both by $f$ and $\mbox{{\sc Spl}}_i^f$. We shall prove that the bipartite distance of $G$ with respect to $\mbox{{\sc Spl}}_i^f$ is at most $\left(2+\frac{k}{50}\right) \eps n^2$ (see Lemma~\ref{cl:splbijcndist}).  Now let us think of generating a bipartition $f_{ij}^{''}$ of $V(G)$ such that, for each $v \in V(G)\setminus X_i$, if we determine $f_{ij}^{''} (v)$ by the same rule used by  our algorithm to determine $f_{ij}(z)$, for each $z \in  V(Z) \setminus X_i$. Note that our algorithm does not find $f_{ij}^{''}$ explicitly, it is used only for the analysis purpose. The number of heavy vertices, with respect to the bipartition $f$, that have different mappings by $f$ and {$f_{ij}''$}, is at most $o(k\eps n)$  with constant probability. So, with a constant probability, $f_{ij}^{''}$ is a {\sc special} bipartition with respect to $f$ by $f_{ij}$.  Note that, if we take $\size{Z}=\Oh(\frac{1}{k^5\eps^2} \log \frac{1}{k\eps})$ many random vertex pairs and  determine the fraction $\chi_{ij}^f$ of pairs that 
form monochromatic edges with respect to the {\sc special} bipartition $f_{ij}^{''}$, we can show that $\chi_{ij}^f \leq  (2 + \frac{k}{20})\eps$, with probability at least $1-2^{-\Omega(\frac{1}{k^3\eps} \log \frac{1}{k \eps})}\geq \frac{9}{10}$. However, we are not finding either $f_{ij}^{''}$ or $\chi_{ij}^f$ explicitly. We just find $\zeta_{ij}$, that is, the fraction of vertex pairs in $Z$ that are monochromatic edges with respect to $f_{ij}'$. But the above argument still holds, since $Z$ is chosen randomly and there exists a $f_{ij}^{''}$, such that  $f_{ij}^{''}(z)=f_{ij}'(z)$, for each $z \in V(Z)$, and the probability distribution of $\zeta_{ij}$ is identical to that of $\chi_{ij}^f$.

\subsection{Soundness}

Let us now consider the case when the bipartite distance of $G$ is at least $(2+ k)\eps n^2$, and $f$ be any bipartition of $V(G)$. To prove the soundness of our algorithm, we introduce the notion of \emph{{\sc derived} bipartition} $\mbox{{\sc Der}}_i^f:V(G)\rightarrow \{L,R\}$ with respect to $f$ by $f_{ij}$ (see Definition~\ref{defi:derivedbipart}), such that $f(v)$, $f_{ij}(v)$ and  $\mbox{{\sc Der}}_i^f(v)$ are identical for each $v \in X_i$. Observe that the bipartite distance of $G$ with respect to any {\sc derived} bipartition is at least $(2+k)\eps n^2$ as well. Similar to the discussion of the completeness, if we generate a bipartition $f_{ij}^{''}$ of $V(G)$, $f_{ij}^{''}$ will be a {\sc derived} bipartition, with respect to $f$ by $f_{ij}$.  If we take $\size{Z}=\Oh(\frac{1}{k^5\eps^2} \log \frac{1}{k\eps})$ many random pairs of vertices and determine the fraction $\chi_{ij}^f$ of pairs that form monochromatic edges with respect to the {\sc derived} bipartition $f_{ij}^{''}$,  we can prove that $\chi_{ij}^f \leq \left(2+\frac{k}{20}\right)\eps$ holds, with probability at most $2^{-\Omega(\frac{1}{k^3\eps} \log \frac{1}{k \eps})}$. We want to re-emphasize that we are not determining $f_{ij}^{''}$, as well as $\chi_{ij}^f$ explicitly. The argument follows due to the facts that $Z$ is chosen randomly and there exists an $f_{ij}^{''}$ such that  $f_{ij}'(z)=f_{ij}^{''}(z)$, for each $z \in V(Z)$, and the probability distribution of $\zeta_{ij}$ is identical to that of $\chi_{ij}^f$. Using the union bound, we can say that 
the algorithm rejects 
with probability at least $\frac{9}{10}$.

%the following holds:$$\mbox{for all} \ i \in [t] and all f_{ij} \in \cF_i, \zeta_{ij}\leq \left(2+\frac{k}{20}\right)\eps $$

\remove{
\paragraph*{Disclaimer} 

We have presented  high level discussions of completeness  and soundness proofs. The actual proofs require more delicate analysis. The formal statements and proofs of completeness and soundness are presented in Section~\ref{sec:correctness}.

such that and $f$  we now check the fraction of edges of $Z$ that have same labels. If this fraction is at most $9 \eps$, we accept, and output that the $G$ is $\eps$-close to being bipartite. If the above condition does not hold for all bipartitions of all $X_i$, with $i\in [t]$, we reject and decide that $G$ is $10 \eps$-far from being bipartite.

between by performing appropriate adjacency queries as well as all the edges with one endpoint in $X_i$ and another endpoint in one vertex in any pair of $Z$. In {\bf Step-2}, 
}

%\subsection{Related work}

%\subsection{Organization of the paper}

%\subsection{Notation}
%For us, $\Oh_{\eps}(1)$ represents $ \Oh(\eps^c)$ for some suitable constant $c$.\\
%$a=(1 \pm \eps)b$ means $(1-\eps)b \leq a \leq (1+\eps)b$.

%\input{intro.tex}
%\input{prelim.tex}
%\input{overview.tex}
\section{Algorithm for Tolerant Bipartite Testing (Proof of Theorem~\ref{theo:mainoverview_algo})}\label{sec:algosec}

%\textcolor{blue}{Need to add few lines.}

In this section, we formalize the ideas discussed in Section~\ref{sec:pfoverview}, and prove Theorem~\ref{theo:mainoverview_algo}.

\subsection*{Formal description of algorithm \tolbip}\label{sec:algo_formal}
\begin{description}
\item[Step-1]
%({\bf Generating a random induced subgraph})
Let $C_1 ,C_2, C_3$ be three suitably chosen large constants and $t:=\lceil{\log \frac{C_1}{k\eps}}\rceil$.
%~\footnote{Here $\widetilde{O}(.)$ hides a factor of poly($\log \frac{1}{\eps}$)}.
\begin{description}
    \item[(i)] We start by generating $t$ many subset of vertices $X_1,\ldots, X_t \subset V(G)$, each with $\lceil{\frac{C_2}{k^3\eps} \log \frac{1}{k\eps}}\rceil$ many vertices, sampled randomly without replacement~\footnote{Since we are assuming $n$ is sufficiently large with respect to $\frac{1}{\eps}$, sampling with and without replacement are the same.}.
    
    %\remove{~\footnote{\comments{Note that since $\eps = o(\sqrt{n})$ where $n$ is the number of vertices, sampling with and without replacement are equivalent for all practical purposes. Otherwise, if $n =\Oh\left(\frac{1}{\eps^2}\right)$, there is a straight forward algorithm with sample complexity $\Oh\left(\frac{1}{\eps^2}\right)$ and query complexity $\Oh\left(\frac{1}{\eps^4}\right)$.}\complain{We will change it later.}}}.
    
    \item[(ii)] We sample $\lceil{\frac{C_3}{k^5\eps^2} \log \frac{1}{k\eps}}\rceil$ many random pairs of vertices, with replacement, and denote those sampled pairs of vertices as $Z$. Note that $X_1,\ldots, X_t, Z$ are generated independent of each other.
    
    \item[(iii)] We  find all the edges with one endpoint in $\cC = X_1 \cup X_2\cup \ldots X_t$ and the other endpoint in one of the vertices of $V(Z)$~\footnote{Recall that $V(Z)$ denotes the set of vertices present in at least one pair in $Z$.}, by performing $\Oh \left(\frac{1}{k^8 \eps^3} \log^2 \frac{1}{k\eps}\right)$ many adjacency queries.
\end{description}

%We first check if $Z$ is disjoint from all $X_i$ s. If not, we REJECT. \textcolor{blue}{Otherwise, we work on the subgraph induced by $\cC$, without making any query further, described as follows.}

\item[Step-2]
%({\bf Subgraph?? Testing})

\begin{description}
\item[(i)] Let $\{a_1,b_1\}, \ldots \{a_{\lambda},b_{\lambda}\}$ be the pairs of vertices of $Z$, where $\lambda = \lceil{\frac{C_3}{k^5\eps^2} \log \frac{1}{k\eps}}\rceil$. Now we find the pairs of $Z$ that are edges in $G$, by performing adjacency queries to all the pairs of vertices of $Z$
(after this step, the algorithm does not make any query further).
%Now we compute the fraction of the above $k$ pairs that form edges in the graph by performing adjacency queries to $G$.

\item[(ii)] For each $i \in [t]$, we do the following:
\begin{description}
    \item[(a)] Let $\cF_i$  denote the set of all possible bipartitions of $X_i$, that is, 
    $$\cF_i=\left \lbrace f_{ij}:X_i\rightarrow \{L,R\}:j \in \left[2^{\size{X_i}-2}\right] \right \rbrace.$$ 
    %Also, for each $j\in [\size{\cB_i}]$, $S_{i}^j$ and $T_{i}^j$ form a partition of $X_i$.

      \item[(b)]For each bipartition $f_{ij}$ (of $X_i$) in $\cF_i$, we extend $f_{ij}$ to $f_{ij}':X_i \cup Z\rightarrow \{L,R\}$ to be a bipartition of $X_i \cup Z$, such that the mapping of each vertex of $X_i$ are identical in $f_{ij}$ and $f_{ij}'$, and is defined as follows:
      
\[  f'_{ij}(z)= \left\{
\begin{array}{ll}
      f_{ij}(z), & z \in X_i\\
      \vspace{3pt}
      L, & z \notin X_i \ \mbox{and} \ \size{N(z) \cap f_{ij}^{-1}(R)} > \size{N(z) \cap f_{ij}^{-1}(L)}+ \frac{k\eps |X_i|}{225000}\\
      \vspace{3pt}
       R, & z \notin X_i \ \mbox{and} \ \size{N(z) \cap f_{ij}^{-1}(L)} > \size{N(z) \cap f_{ij}^{-1}(R)}+\frac{k\eps |X_i|}{225000}\\
      \vspace{3pt}
     {\mbox{L or $R$ arbitrarily}} , & \mbox{otherwise} \\
\end{array} 
\right. \]

%For each vertex $z \in V(Z) \setminus X_i$, $f_{ij}'$ maps $z$ to $L$ or $R$ in a \emph{specific} way. The full description of $f_{ij}':Z \cup X_i \rightarrow \{L,R\}$ is as follows:

Note that this step can be performed from the adjacency information between the vertices of $\cC$ and $Z$, which have already been computed before.

%by only seeing the edges between $\cC$ and $Z$ which have already been computed earlier.
%\remove{Recall that $f_{ij}$ is a bipartition of $X_i$, as defined in {\bf Step-2(ii)(a)}. Now we construct a new bipartition $f'_{ij}$ by extending $f_{ij}$ (over $X_i$) to $X_i \cup Z$, and thereafter deciding the labels of the vertices in $Z \cup X_i$.}

    \item[(c)]
      We now find the fraction of the vertex pairs of $Z$ that are edges and have the same label with respect to $f_{ij}'$, that is,
   $$ \zeta_{ij}=2 \cdot \frac{\size{\left \lbrace \{a_\ell,b_\ell\}: \ell \in [\lambda], \{a_\ell,b_\ell\}\in E(G)~\mbox{and}~f'_{ij}(a_\ell)=f'_{ij}(b_\ell)\right \rbrace}}{\lambda}~\footnote{$2$ is multiplied, as in the definition of $d_{bip}(G,f)$, each edge $\{u,v\} \in E(G)$ with $f(u)=f(v)$ is counted twice.}.$$
   %Note that this step can be performed by only seeing the induced subgraph $\cC$.\textcolor{green}{Rahul: seems unnecessary}
       
    %We now find the fraction of vertex pairs of $Z$ that form edges and both the vertices in the pairs have same $f_{ij}'$ values (it may be either $L$ or $R$)   
       
   \item[(d)] If $\zeta_{ij} \leq \left(2+\frac{k}{20}\right)\eps$, we {\sc Accept} $G$ as $\eps$-close to being bipartite,  and {\sc Quit} the algorithm.
    
\end{description}    

    \item[(iii)] If we arrive at this step, then $\zeta_{ij} > \left(2+\frac{k}{20}\right)\eps$, for each $i \in [t]$ and $f_{ij} \in \cF_i$ in Step-{ (ii)}. We {\sc Reject} and declare that $G$ is $(2 + k)\eps$-far from being bipartite. 
    
    %()

\end{description}

\end{description}

We split the analysis of algorithm \tolbip into five parts:
\begin{description}

\item[Completeness:] If $G$ is $\eps$-close to being bipartite, then   \tolbip reports the same, with probability at least $\frac{9}{10}$.

\item[Soundness:] If $G$ is $(2+k) \eps$-far from being bipartite, then \tolbip reports the same, with probability at least $\frac{9}{10}$.

\item[Sample Complexity:] The sample complexity of \tolbip is $\Oh(\frac{1}{k^5\eps^2} \log \frac{1}{k\eps})$.

\item[Query Complexity:] The query complexity of \tolbip is $\Oh (\frac{1}{k^8\eps^3} \log^2 \frac{1}{k\eps})$.

\item[Time Complexity:] The time complexity of \tolbip is $2^{\Oh(\frac{1}{k^3\eps} \log \frac{1}{k\eps})}$.

\end{description}

The last three quantities can be computed from the description of \tolbip. In {\bf Step-1(i)}, we sample vertices of $G$ to generate $t=\lceil{\log \frac{C_1}{k\eps}}\rceil$ subsets, each with $\lceil{\frac{C_2}{k^3\eps} \log \frac{1}{k\eps}}\rceil$ many vertices. Thereafter in {\bf Step-1(ii)} and {\bf Step-1(iii)}, we randomly choose $\lceil{\frac{C_3}{k^5\eps^2} \log \frac{1}{k\eps}}\rceil$ many pairs of vertices and perform adjacency queries for each vertex in any pair of $Z$ to every $X_i$. Thus the sample complexity of \tolbip is $\Oh(\frac{1}{k^5\eps^2} \log \frac{1}{k\eps})$ and query complexity is $\Oh(\frac{1}{k^8\eps^3} \log^2 \frac{1}{k\eps})$. The time complexity of the algorithm is $2^{\Oh(\frac{1}{k^3\eps} \log \frac{1}{k\eps})}$, which follows from {\bf Step-2(ii)}, that dominates the running time.

%Note that each such subgraph can be generated by using $\Oh\left(\frac{1}{\eps^4}\right)$ queries, and hence the claim about query complexity follows.

\section{Proof of Correctness of {\bf \texorpdfstring{\tolbip}{\space}}}\label{sec:correctness}

Before proceeding to the proof, we introduce some definitions for classifying the vertices of the graph, with respect to any particular bipartition, into two categories: $(i)$ \emph{\textbf{heavy}} vertices, and $(ii)$ \emph{\textbf{balanced}} vertices. These definitions will be mostly used in the proof of completeness. Informally speaking, a vertex $v$ is said to be \textbf{heavy} with respect to a bipartition $f$, if it has \emph{substantially} large number of neighbors in one side of the bipartition (either $L$ or $R$), as compared to the other side.

\begin{defi}[Heavy vertex]\label{defi:heavyvertex}
A vertex $v \in V$ is said to be \emph{{\bf $L$-heavy} with respect to a bipartition $f$}, if it satisfies two conditions:

\begin{itemize}
    \item[(i)] $\size{N(v) \cap f^{-1}(L)}\geq \size{N(v) \cap f^{-1}(R)} + \frac{k \eps n}{150}$;
    
    \
    \item[(ii)] If $\size{N(v) \cap f^{-1}(R)}\geq \frac{1}{(1 + \frac{k}{200})}\frac{k \eps n}{150}$, then $$\mbox{$\size{N(v) \cap f^{-1}(L)} \geq (1 + \frac{k}{200})\size{N(v) \cap  f^{-1}(R)}$};$$ 
    \end{itemize}

We define {\bf $R$-heavy} vertices analogously. The union of the set of {\bf $L$-heavy} and {\bf $R$-heavy} vertices, with respect to a bipartition $f$, is defined to be the set of {heavy} vertices (with respect to $f$), and is denoted by $\cH_f$.

Similarly, a vertex $v$ is said to be \textbf{balanced} if the number of neighbors of $v$ are \emph{similar} in both $L$ and $R$, with respect to a bipartition $f$. We define it formally as follows:

%Note that, for such a vertex $v$, $\mbox{{\sc Spl}}_i(v)= R$.

%All vertices that are not $L$-heavy and $R$-heavy are defined as {\bf Balanced Type $1$} with respect to $f$ and denoted as $\cB$.

%%If $\size{N(v) \cap f_{ij}^{-1}(L)}$ and $\size{N(v) \cap f_{ij}^{-1}(R)}$ are large, then $d_A(v) \geq 3d_B(v)$.

\end{defi}
%\textcolor{red}{Rahul: Why mention SPL here?}

\begin{defi}[Balanced vertex]\label{defi:balancedvertex}
A vertex $v \in V$ is said to be \emph{{\bf balanced} with respect to a bipartition $f$}, if $v\notin \cH_f$, that is, it satisfies at least one of the following conditions: 

\begin{description}
    \item[(i) Type 1:] $\size{\size{N(v) \cap f^{-1}(R)} -  \size{N(v) \cap f^{-1}(L)}}< \frac{k\eps n}{150}$;

\
   \item[(ii) Type 2:] 
        Either 
        $$
            \size{N(v)\cap f^{-1}(L)} \leq \size{N(v)\cap f^{-1}(R)} < {(1+\frac{k}{200})} \size{N(v)\cap f^{-1}(L)},
        $$
        or,
        $$
            \size{N(v)\cap f^{-1}(R)} \leq \size{N(v)\cap f^{-1}(L)} < {(1+\frac{k}{200})} \size{N(v)\cap f^{-1}(R)}.
        $$

    %If $\size{N(v) \cap f_{ij}^{-1}(L)}\geq \frac{\eps n}{10}$ and $\size{N(v) \cap f_{ij}^{-1}(R)}\geq \frac{\eps n}{30}$ , then $\size{N(v) \cap f_{ij}^{-1}(L)} \geq 3 \size{N(v) \cap f_{ij}^{-1}(R)}$.
\end{description}

The set of {balanced vertices of type 1} with respect to $f$ is denoted as $\cB_f^1$, and the set of {balanced vertices of type 2} with respect to $f$ is denoted as $\cB_f^2$. The union of $\cB_f^1$ and $\cB_f^2$ is denoted by $\cB_f.$ {Note that $\cB_f^1$ and $\cB_f^2$ may not be disjoint.}
\end{defi}

In order to prove the completeness (in Section~\ref{sec:correctness1}), we also use a notion of \emph{{\sc {\sc  {\sc  special}}} bipartition} to be defined below. The definition of {\sc  special} bipartition is based on an optimal bipartition $f$ of $V(G)$, and notions of heavy and balanced vertices. We would also like to note that, later in Lemma~\ref{cl:splbijcndist}, we show that when $d_{bip}(G)\leq \eps n^2$, the bipartite distance of $G$ with respect to any {\sc  special} bipartition is bounded by $(2+ \frac{k}{50})\eps n^2$. 

\begin{defi}[{\sc  special} bipartition]\label{defi:splbijection}
Let $d_{bip}(G) \leq \eps n^2$, and $f:V(G)\rightarrow \{L,R\}$ be an optimal bipartition of $V(G)$, that is, $d_{bip}(G,f)\leq \eps n^2$, and there does not exist any bipartition $g$ such that $d_{bip}(G,g) < d_{bip}(G,f)$. For an $X_i$ selected in {\bf Step-1(i)} of the algorithm, let $f_{ij} \in \cF_i$ be the bipartition of $X_i$ such that $f\mid_{X_i}=f_{ij}$. Then bipartition $\mbox{{\sc Spl}}_i^f:V(G)\rightarrow \{L,R\}$ is said to be a {\bf {\sc special} bipartition} with respect to $f$ by $f_{ij}$ such that
%\textcolor{green}{ (Rahul: I think last three lines should be removed and last line should be " Then, a {\bf {\sc  special} bipartition} with respect to $X_i$,  denoted by $\mbox{{\sc Spl}}_i: V(G)\rightarrow \{L,R\}$ represents any bipartition of $V(G)$ such that") {\sc  special} does not need a superscript? Cant we start with come fixed optimal bipartition?  {\sc  special} is a family of bipartitions? So we could ask if $f'_{ij}$ is in the set $SPL_i$?}

\begin{itemize}
\item $\mbox{{\sc Spl}}_i^f\mid_{X_i}=f \mid _{X_i}=f_{ij} $;
     
\item There exists a subset $\cH_f' \subset \cH_f$ such that $\size{\cH_f'}\geq (1-o(k\eps))\size{\cH_f}$, and for each $v \in \cH_f'$, $\mbox{{\sc Spl}}_i^f(v)$ is defined as follows:
      \[  \mbox{{\sc Spl}}_i^f(v)= \left\{
\begin{array}{ll}
     
      R, & v \notin X_i \ \mbox{and} \ v~\mbox{is}~ L-\mbox{heavy} \\
      \vspace{3pt}
            L, & v \notin X_i \ \mbox{and} \ v~\mbox{is}~ R-\mbox{heavy}  
\end{array} 
\right. \]

\item For each $v \notin (\cH_f' \cup X_i)$, $\mbox{{\sc Spl}}_i^f(v)$ is set to $L$ or $R$ arbitrarily.

%\textcolor{green}{arbitrarily or randomly???? If arbitrarily, all balanced vertices can be put on $L$-side. Will this cause a problem? Will doing this affect Lemma 4.9?\newline  Another thing, Fischer et. al. need randomness in their labelling function, that is why they need to deal with random restrictions??, but if we are allowed to arbitrarily map balanced vertices, our labelling function is completely determined given $X_i$.}
     
\end{itemize}

\end{defi}

In our proof of the soundness theorem (in Section~\ref{sec:correctness2}), we need the notion of \emph{{\sc derived} bipartition}. Unlike the definition of {\sc  special} bipartition, the definition of {\sc derived} bipartition is more general, in the sense that it is not defined based on either any optimal bipartition, or on heavy or balanced vertices.

\begin{defi}[{\sc derived} bipartition]\label{defi:derivedbipart}
Let $f:V(G)\rightarrow \{L,R\}$ be a bipartition of $V(G)$. For an $X_i$ selected in {\bf Step-1(i)} of the algorithm, let $f_{ij} \in \cF_i$ be the bipartition of $X_i$ such that $f\mid_{X_i}=f_{ij}$. A bipartition   $\mbox{{\sc Der}}_i^f:V(G)\rightarrow \{L,R\}$ is said to be  {\bf \emph{{\sc derived}}} bipartition with respect to $f$ by $f_{ij}$, if $\mbox{{\sc Der}}_i^f\mid_{X_i}=f\mid_{X_i}=f_{ij}$.

% \begin{itemize}
% \item  $\mbox{{\sc Der}}_i\mid_{X_i}=f\mid_{X_i}=f_{ij}$;
%     \item There exists a subset $\cH_f' \subset \cH_f$ such that $\size{\cH_f'}\geq (1-o(k\eps))\size{\cH_f}$ , and for each $v \in \cH_f'$
%      \[  \mbox{{\sc Der}}_i(v)= \left\{
%\begin{array}{ll}
%     
%      R, & v \notin X_i \ and \ v~\mbox{is}~ L-\mbox{heavy} \\
%      \vspace{3pt}
%            L, & v \notin X_i \ and \ v~\mbox{is}~ R-\mbox{heavy}  
%\end{array} 
%\right. \]
%\item For each $v \notin (\cH_f' \cup X_i)$, $\mbox{{\sc Der}}_i(v)$ is set to $L$ or $R$ arbitrarily.
     
%\end{itemize}

% \[  \mbox{{\sc Der}}_i(v)= \left\{
%\begin{array}{ll}
%      f_{ij}(v), & v \in X_i \\
%      \vspace{3pt}
%      L, & v \notin X_i \ and \ v~\mbox{is}~ L-\mbox{heavy} \\
%      \vspace{3pt}
%            R, & v \notin X_i \ and \ v~\mbox{is}~ R-\mbox{heavy}  \\
%      \vspace{3pt}
%      \mbox{L or R arbitrarily}, & \mbox{Otherwise} \\
%\end{array} 
%\right. \]
\end{defi}
%\textcolor{green}{(Rahul: shouldn't the above be $DER^{f_{ij}}_i$?.)}
%\textcolor{green}{(suggested alternative) Let $f_{ij}:X_i\rightarrow \{L,R\}$ be a bipartition of $X_i$, for an $X_i$ selected in {\bf Step-1}. That is, $f_{ij}\in\cF_i$. Then, $\mbox{{\sc Der}}_i^{f_{ij}}:V(G)\rightarrow \{L,R\}$ denotes any bipartition of $V(G)$ such that $\mbox{{\sc Der}}_i^{f_{ij}}\mid_{X_i}=f_{ij}$. We call such a bipartition \emph{a {\bf \emph{derived} bipartition} of $f_{ij}$}. (given definition seems weird to me).\newline}
%\newline 
%We now extend the definition of derived partition to the case when $G$ is $\eps$-close to being bipartite. 

\subsection{Proof of Completeness}\label{sec:correctness1}

In this section, we prove the following theorem:

\begin{theo}\label{theo:completeness}
Let us assume $G$ is $\eps$-close to being bipartite. Then \tolbip reports the same, with probability at least $\frac{9}{10}$.
\end{theo}
The proof of Theorem~\ref{theo:completeness} will cricially use the following lemma, which says that the bipartite distance of $G$ with respect to any {\sc  special} bipartition is bounded by a $\left(2+\frac{k}{50}\right) \eps n^2$.

\begin{lem}[{\sc  special} bipartition lemma]\label{cl:splbijcndist} 
Let $f$ be a bipartition such that $d_{bip}(G,f) \leq \eps n^2$ and there does not exist any bipartition $g$ such that $d_{bip}(G,g) < d_{bip}(G,f)$. 
For any {\sc  special} bipartition $\mbox{{\sc Spl}}_i^f$ with respect to $f$, $d_{bip}(G, \mbox{{\sc Spl}}_i^f) \leq \left(2+\frac{k}{50}\right) \eps n^2$.
%is not large 
%Total distance due to {\emph heavy} vertices $\leq 2 \eps n^2$.
\end{lem}

We will prove the above lemma later. For now, we want to establish (in Lemma~\ref{lem:totaldist}) that there exists an $i \in [t]$ and a$f_{ij} \in \cF_i$ which can be thought of as a \emph{random restriction} of some {\sc  special} bipartition with respect to $f$ by $f_{ij}$. In other words, Lemma~\ref{lem:totaldist} basically states that if $G$ is $\epsilon$-close to being bipartite, then the extension  according to the rule in {\bf Step-2(ii)(b)} of the mapping  obtained by restricting an optimal bipartition to a random $X_i$  is likely to correspond to a {\sc  special} bipartition, and therefore, the number of monochromatic edges (with respect to a {\sc  special} bipartition) in the randomly picked $Z$ is likely to be low with respect to that bipartition. Thus, $\zeta_{ij}$ must be low for some $i,j$ with high probability.
 
To prove Lemma~\ref{lem:totaldist},  we need the following lemma (Lemma~\ref{cl:heavyvertex}) about heavy vertices. In Lemma~\ref{cl:heavyvertex}, we basically prove that a heavy vertex with respect to a bipartition $f$ will have significantly more neighbors in the part of $X_i$, that corresponds to the heavy side of that vertex (with respect to $f$). Basically, if a  vertex $v$ is L-heavy with respect to $f$, it has more neighbors in the subset of $X_i$ on the L-side as compared to the subset of $X_i$ on the R-side of $f$. 
 
\begin{lem}[Heavy vertex lemma]\label{cl:heavyvertex}
Let $f$ be a bipartition of $G$. Consider a vertex $v\in V$.
\begin{description}
    \item[(i)] 
    For every L-heavy vertex $v$, $\size{N(v)\cap f^{-1}(L) \cap X_i} - \size{N(v)\cap f^{-1}(R) \cap X_i} \geq {\frac{k^2 \eps \size{X_i}}{225000}}$ with probability at least $1 - o(k\eps)$.
    \item[(ii)] For every R-heavy vertex $v$, $\size{N(v)\cap f^{-1}(L) \cap X_i} - \size{N(v)\cap f^{-1}(R) \cap X_i} \geq {\frac{ k^2 \eps \size{X_i}}{225000}}$ with probability at least $1 - o(k\eps)$.
\end{description}
\end{lem}

We would like to note that Lemma~\ref{cl:heavyvertex}  holds for any bipartition. However, we will use it only for completeness with resepct to  an optimal bipartition $f$.
 
\color{black} 
 
\begin{lem}\label{lem:totaldist}
If $d_{bip}(G)\leq \eps n^2$, then there exists an $i \in [t]$ and $f_{ij}\in \cF_i$ such that $\zeta_{ij} \leq \left(2+\frac{k}{20}\right) \eps$ holds, with probability at least $1-o(k\eps)$.
\end{lem}

\begin{proof}
Let $f$ be an optimal bipartition such that $d_{bip}(G,f)\leq \eps n^2$. First, consider a {\sc  special} bipartition $\mbox{{\sc Spl}}_i^f$, and consider a set of random vertex pairs $Y$ such that $|Y|=|Z|$. Now consider the fraction of monochromatic edges of $Y$, with respect to the bipartition $\mbox{{\sc Spl}}_i^f$, that is,
$$\chi^f_{ij} = 2 \cdot \frac{\size{\left\lbrace \{a,b\}\in Y: \{a,b\}\in E(G)~\mbox{and}~\mbox{{\sc Spl}}_i^f(a)=\mbox{{\sc Spl}}_i^f(b)\right \rbrace}}{\size{Y}}.$$

\begin{obs}\label{obs:obs1}
$\chi^f_{ij} \leq \left(2+\frac{k}{20}\right) \eps$ holds, with probability at least $\frac{9}{10}$.
\end{obs}

\begin{proof}
By Lemma~\ref{cl:splbijcndist}, we know that when $d_{bip}(G) \leq \eps n^2$, $d_{bip}(G,\mbox{{\sc Spl}}_i^f)\leq \left(2+\frac{k}{50}\right)\eps n^2$. So, $\E[\chi_{ij}^f]\leq \left(2+\frac{k}{50}\right)\eps$. Using Chernoff bound (see Lemma~\ref{lem:cher_bound3}), we can say that
$$\mbox{$\pr (\chi^f_{ij} \geq \left(2+\frac{k}{20}\right) \eps) \leq \frac{1}{2^{\Omega(\frac{1}{k^3\eps} \log \frac{1}{k\eps})}}\leq \frac{1}{10}.$}$$ %So, we are done with the proof of Observation~\ref{obs:obs1}.\textcolor{green}{remove last line}

%\textcolor{blue}{TO CHECK THE BOUND}

\end{proof}

Now, we claim that bounding $\chi^f_{ij} $ is equivalent to bounding $\zeta_{ij}$.

\begin{cl}\label{cl:zetachi}
For any $i \in [t]$, there exists a bipartition $f_{ij}\in \cF_i$ such that the probability distribution of $\zeta_{ij}$ is identical to that of $\chi^f_{ij}$, for some {\sc special} bipartition $f$ with respect to $f_{ij}$, with probability at least $\frac{1}{2}$.
\end{cl}

As $t =\Oh(\log \frac{1}{k\eps})$, the above claim implies that there exists an $i \in [t]$ and $f_{ij}\in \cF_i$ such that the probability distribution of $\zeta_{ij}$ is identical to that of $\chi^f_{ij}$, with probability at least $1-o(k\eps)$.

Now we prove Claim~\ref{cl:zetachi}.  Recall the procedure of determining $\zeta_{ij}$ as described in {\bf Step 2} of algorithm \tolbip presented in Section~\ref{sec:algosec}. 
\begin{description}
  \item[Fact 1:]   
  For any vertex $v \in \cH_f  \cap Z$, $\mbox{{\sc Spl}}_i^f(v)=f_{ij}'(v)$, with probability at least $1-o(k\eps)$, where $\cH_f$ denotes the set of heavy vertices of $X_i$ with respect to the bipartition $f$. This follows according to Claim~\ref{cl:heavyvertex}, along with the definition of $f_{ij}'(z)$.
  
  %Because of the way we determine $f_{ij}'(z)$ for each $z  \in Z$, by Lemma~\ref{cl:largeapprox}, each vertex in $\cH_f  \cap Z$, $\mbox{{\sc Spl}}_i^f(v)=f_{ij}'(v)$ with probability at least $1-o(k\eps)$. 

  \item[Fact 2:] 
  Consider a bipartition $f_{ij} \in \cF_i$ of $X_i$, and its extension $f^{'}_{ij}$ to $X_i\cup Z$, as considered in the algorithm. 
  Assume a bipartition $f_{ij}^{''}$ of $V(G)$, constructed by extending $f'_{ij}$ according to the rule of {\bf Step-2(ii)(b)} of the algorithm.
  From Heavy vertex lemma (Lemma~\ref{cl:heavyvertex}), we know that the expected number of vertices in $\cH_f$ such that  $f_{ij}^{''}(v)\neq f(v)$, is at most $o(k\eps)\size{\cH_f}$. Using Markov inequality, we can say that, with probability at least $\frac{1}{2}$, the number of vertices in $\cH_f $ such that  $f_{ij}^{''}(v)\neq f(v)$, is at most $o(k\eps)\size{\cH_f}$. Thus, with probability at least $\frac{1}{2}$, there exists a set of vertices $\cH_f'$ such that $f_{ij}^{''}(v)= f(v)$ holds for at least $(1-o(k\eps))\size{\cH_f'}$ vertices. Note that the bipartition $f_{ij}^{''}$ is a {\sc  special} bipartition $f$ with respect to $f_{ij}$.

  %\color{blue}
  
   %Recall that $f$ is an optimal bipartition of $V(G)$. Consider bipartition $f_{ij} \in \cF_i$ of $X_i$, and bipartition $f^{'}_{ij}$ of $X_i\cup Z$ to $\{L,R\}$, as considered in the algorithm. For the sake of argument, let us extend the domain of $f_{ij}^{'}$ to $V(G)$, by the rule (in {\bf Step-2(ii)(b)}) that is used to map the vertices of $X_i \cup Z$ for constructing $f_{ij}'$. We refer to this bipartition of $V(G)$ as $f_{ij}^{''}$. Now consider the set $\cH_f$ of heavy vertices with respect to $f$.  By Lemma~\ref{cl:heavyvertex},  the expected number of vertices in $\cH_f$ such that  $f_{ij}^{''}(v)\neq f(v)$, is at most $o(k \eps)\size{\cH_f}$.  By Markov Inequality, the  number of vertices in $\cH_f $ such that  $f_{ij}^{''}(v)\neq f(v)$, is at most $o(k \eps)\size{\cH_f}$, with probability at least $1/2$. So there exists $\cH_f'$ such that $f_{ij}^{''}(v)= f(v)$ holds for at least $(1-o(k\eps))\size{\cH_f}$ vertices, with probability at least $\frac{1}{2}$.  Note that such $f_{ij}^{''}$ is a {\sc  special} bipartition of $f$ with respect to $f_{ij}$.
   
   \color{black}
   
   %\remove{Let us think of the following instance (for argument purpose only). Let $f_{ij}^{''}$ is a bipartition from $V(G)$ to $\{L,R\}$, where for each prove Claim~\ref{cl:zetachi}.  Recall the procedure of determining $\zeta_{ij}$ as described in {\bf Step 2} of our algorithm presented in Section $v \in V(G)$, $f_{ij}^{''}(v) $ is determined by the same rule that we are using to determine $f_{ij}^{''}(z)$ for each $z \in Z$. {Replace last few lines with: Let $f$ be an optimal bipartition of $V(G)$.}}
   
  \end{description}

  From {\bf Fact 1} and {\bf Fact 2}, we can deduce that, there exists a {\sc  special} bipartition $\mbox{{\sc Spl}}_i^f$ such that $\mbox{{\sc Spl}}_i^f(v)=f_{ij}'(v)$ for each $z \in Z$.  
 
 Since we choose $Z$ uniformly at random, Lemma~\ref{lem:totaldist} follows.
 
 %Hence, the lemma (Lemma~\ref{lem:totaldist}) follows as we choose $Z$ uniformly at random.
\end{proof}
% \textcolor{green}{Please demarcate clearly where proof of lemma 4.8 ends.}

According to the description of algorithm \tolbip, the algorithm reports that $d_{bip}(G)\leq \eps n^2$, if there exists a $\zeta_{ij}$ such that $\zeta_{ij}\leq \left(2+\frac{k}{20}\right)\eps$, for some $i \in [t]$ and $j \in [2^{\size{X_i}-2}]$. Hence, by Lemma~\ref{lem:totaldist}, we are done with the proof of the completeness theorem (Theorem~\ref{theo:completeness}).

Now we focus on proving {\sc  special} bipartition lemma (Lemma~\ref{cl:splbijcndist}) and Heavy vertex lemma (Lemma~\ref{cl:heavyvertex}), starting with the proof of {\sc  special} bipartition lemma.

\subsection*{Proof of {\sc  special} bipartition lemma (Lemma~\ref{cl:splbijcndist})}
%\complain{Need to change: The next lemma is the key technical argument of the completeness proof. Informally, it states that starting from an optimal bipartition, as long as most of the heavy vertices are placed correctly, moving around the rest of the vertices will not increase the bipartite distance by a lot.  \comments{This text to be inserted suitably.}}

The idea of the proof relies on decomposing the bipartite distance with respect to a {\sc special} bipartition into a sum of three terms and then carefully bounding the cost of each of those parts individually.

%Observe that $d_{bip}(G,\mbox{{\sc Spl}}_i^f)$ can be expressed as

Let us first recall the definition of bipartite distance of $G$ with respect to a special bipartition $\mbox{{\sc Spl}}_i^f$.
\begin{equation}\label{eqn:splbijection_definition}
  d_{bip}(G,\mbox{{\sc Spl}}_i^f)= \size{\left\{ (u,v) \in E(G):  \mbox{{\sc Spl}}_i^f(u)=\mbox{{\sc Spl}}_i^f(v)\right\}}.  
\end{equation}
{By abuse of notation, here we are denoting $E(G)$ as the set of ordered edges.}

We will upper bound $d_{bip}(G,\mbox{{\sc Spl}}_i^f)$ as the sum of three terms defined below. Here $\cH_f$ and $\cB_f$ denote the set of heavy vertices and balanced vertices (with respect to $f$), as defined in Definition~\ref{defi:heavyvertex} and Definition~\ref{defi:balancedvertex}, respectively. Also, $\cH_f' \subseteq \cH_f$ denotes the set of vertices of $\cH_f$ that are mapped according to $f$, as defined in the definition of {\sc special} bipartition in Definition~\ref{defi:splbijection}. The three terms that are used to upper bound $d_{bip}(G,\mbox{{\sc Spl}}_i^f)$ are as follows:
\begin{description}
    \item[(a)] $D_{\cH_f'  \cup X_i, \cH_f' \cup X_i}=   \size{\left\{ (u,v) \in E(G): u \in \cH_f' \cup X_i~\mbox{and}~v \in \cH_f' \cup X_i, \mbox{{\sc Spl}}_i^f(u)=\mbox{{\sc Spl}}_i^f(v)\right\}}$;
    
    \item[(b)] $D_{\cH_f \setminus (\cH_f' \cup X_i), V(G)}=$   
    
    $~~~~~~~~~\size{\left\{ (u,v)\in E(G) : u \in \cH_f \setminus(\cH_f' \cup X_i), \&~v\in  V(G), \mbox{{\sc Spl}}_i^f(u)=\mbox{{\sc Spl}}_i^f(v)\right\}}$;
    
    \item[(c)] $D_{\cB_f \setminus X_i, V(G)}=   \size{\left\{ (u,v)\in E(G): u \in \cB_f \setminus X_i ~\mbox{and} \ v \in V(G) , \mbox{{\sc Spl}}_i^f(u)=\mbox{{\sc Spl}}_i^f(v)\right\}}$.
\end{description}
    
Now from Equation~\ref{eqn:splbijection_definition} along with the above definitions, we can upper bound $d_{bip}(G,\mbox{{\sc Spl}}_i^f)$ as follows:
\begin{eqnarray}\label{eqn:splbijection}
d_{bip}(G,\mbox{{\sc Spl}}_i^f) \leq D_{\cH_f' \cup X_i, \cH_f' \cup X_i}+ D_{\cH_f \setminus (\cH_f' \cup X_i), V(G)} +{ D_{\cB_f \setminus X_i, V(G)}}.
\end{eqnarray}

We now upper bound $d_{bip}(G,\mbox{{\sc Spl}}_i^f)$ by bounding each term on the right hand side of the above expression separately, via the two following claims which we will prove later.

\begin{cl}\label{cl:boundheavy}
\begin{description}
    \item[(i)] {$D_{\cH_f' \cup X_i, \cH_f' \cup X_i}\leq d_{bip}(G,f)- \Pi$, where \\ 
$$\Pi :=   \left[\sum_{v\in \cB_f \setminus X_i:f(v)=L}|N(v)\cap f^{-1}(L)|+ \sum_{v\in  \cB_f \setminus X_i:f(v)=R}|N(v)\cap f^{-1}(R)|\right];$$}

    \item[(ii)] $D_{\cH_f \setminus (\cH_f' \cup X_i), V(G)} \leq o(k\eps)n^2$;
\end{description}
\end{cl}

\begin{cl}\label{cl:boundbalanced}
$D_{\cB_f \setminus X_i, V(G)} \leq  2\left(1 + \frac{k}{400}\right)\Pi + \frac{ k\eps n^2}{150}$.
\end{cl}

Assuming Claim~\ref{cl:boundheavy} and Claim~\ref{cl:boundbalanced} hold, along with Equation~\ref{eqn:splbijection}, $d_{bip}(G,\mbox{{\sc Spl}}_i^f)$ can be upper bounded as follows:
\begin{eqnarray*}
 d_{bip}(G,\mbox{{\sc Spl}}_i^f) &\leq& d_{bip}(G,f)-\Pi + o(k\eps)n^2 + 2\left(1 + \frac{k}{400}\right) \Pi +  \frac{k\eps n ^2}{150} \\
&\leq& d_{bip}(G,f)+\Pi + \frac{k }{200}\Pi+\frac{k\eps n ^2}{100}.
\end{eqnarray*}

Note that $\Pi \leq d_{bip}(G,f)$ and $d_{bip}(G,f)\leq \eps n^2$. Hence, we can say the following:
$$ {d_{bip}(G,\mbox{{\sc Spl}}_i^f)} \leq \left(2+ \frac{k}{50}\right)\eps n^2.$$

So, we are done with the proof of the {\sc special} bipartition lemma. We are left with the proofs of Claim~\ref{cl:boundheavy} and Claim~\ref{cl:boundbalanced}.

\begin{proof}[{\bf Proof of Claim~\ref{cl:boundheavy}}]
(i) We use the following observation in our proof. The observation follows due to the fact that the bipartition $f$ considered is an optimal bipartition.

\begin{obs}\label{obs:optimalplacement}
Let $v$ be a $L$-heavy vertex $v$ with respect to $f$. Then $f(v) = R$. Similarly, for every R-heavy vertex $v$ with respect to $f$, $f(v) = L$.
\end{obs}

Following the definition of {\sc special} bipartition, we know that there exists a set of vertices $\cH'_f \subset \cH_f$ such that $\size{\cH'_f}\geq (1-o(k\eps))\size{\cH_f}$, and for each $v \in \cH'_f$, the following holds:
      \[  \mbox{{\sc Spl}}_i^f(v)= \left\{
\begin{array}{ll}
     
      R, & v \notin X_i \ \mbox{and} \ v~\mbox{is}~ L-\mbox{heavy} \\
      \vspace{3pt}
            L, & v \notin X_i \ \mbox{and} \ v~\mbox{is}~ R-\mbox{heavy}  
\end{array} 
\right. \]

By Observation~\ref{obs:optimalplacement}, we know that for every $v \in \cH'_f$, $\mbox{{\sc Spl}}_i^f(v)=f(v)$.
Moreover, for each $v \in X_i$, $\mbox{{\sc Spl}}_i^f(v)=f(v)$, following the definition of {\sc special} bipartition $\mbox{{\sc Spl}}_i^f$. Thus for every $v \in \cH'_f \cup X_i$, $\mbox{{\sc Spl}}_i^f(v)=f(v)$. Hence,
\begin{eqnarray*}
&& D_{\cH_f' \cup X_i, \cH_f' \cup X_i} \\ &=& \size{\left\{ (u,v) \in E(G) :u \in \cH_f' \cup X_i~\mbox{and}~v \in  \cH_f' \cup X_i,  \mbox{{\sc Spl}}_i^f(u)=\mbox{{\sc Spl}}_i^f(v)\right\}}\\
&=& \size{\left\{(u,v)\in E(G) : u \in \cH_f' \cup X_i,~\mbox{and} \ v  \in \cH_f' \cup X_i,  f(u)=f(v)\right\}} \\
&&~~~~~~~~~~~~~~~~~~~~~~~~~~~~~~~~~~~~~~~~~~~(\because \mbox{for every $v \in \cH'_f \cup X_i$, $\mbox{{\sc Spl}}_i^f(v)=f(v)$}) \\
%&\leq& \size{(u, v) \in E(G): u \in V(G)~\mbox{and} \ v \in V(G), f(u)=f(v)}\\
&=& d_{bip}(G,f)- \\
&& ~~~~~~~~~~~\left[\sum_{v\in V \setminus (\cH_f' \cup X_i):f(v)=L}|N(v)\cap f^{-1}(L)|+ \sum_{v\in  V \setminus (\cH_f' \cup X_i):f(v)=R}|N(v)\cap f^{-1}(R)|\right] \\
&\leq& d_{bip}(G,f)-  \left[\sum_{v\in \cB_f \setminus X_i:f(v)=L}|N(v)\cap f^{-1}(L)|+ \sum_{v\in  \cB_f \setminus X_i:f(v)=R}|N(v)\cap f^{-1}(R)|\right] \\
&=& d_{bip}(G,f)-\Pi.
\end{eqnarray*}
\color{black}

(ii) By the definition of $\cH_f'$, we know that $\size{\cH_f \setminus (\cH_f' \cup X_i)}$ is upper bounded by $o(k\eps)\size{\cH_f}$. Following the definition of
$D_{\cH_f \setminus (\cH_f' \cup X_i), V(G)}$, we can say the following: 
\begin{eqnarray*}
&& D_{\cH_f \setminus (\cH_f' \cup X_i), V(G)}\\ &=&  \size{\left\{(u,v) \in E(G): u \in \cH_f \setminus(\cH_f' \cup X_i)~\mbox{and} \ v\in V(G), \mbox{{\sc Spl}}_i^f(u)=\mbox{{\sc Spl}}_i^f(v)\right\}} \\
&\leq&  \size{\cH_f \setminus (\cH_f' \cup X_i)}\times \size{V(G)}
= o(k\eps) \size{\cH_f} \times n 
\leq o(k\eps)n^2.
\end{eqnarray*}
The last inequality follows as  $\size{\cH_f}$ is at most $n$.
%
%Note that from Claim~\ref{}, we know that the probability that a heavy vertex is wrongly placed is $\leq o(\eps)$.
%
%Since we are choosing the vertices of $X_i$ randomly, the expected number of wrongly placed vertices $\leq o(\eps)n$.
%
%Thus we can also say that with probability $1 - \Oh_{\eps'}(1)$~\footnote{Note that $\eps' < \eps$ chosen suitably.}, the total number of wrongly placed heavy vertices is at most $o(\eps) n$.
%
%So $D_{\cH_f \setminus (\cH_f' \cup X_i), V(G)} \leq o(\eps)n^2$.
%
%
%
\end{proof}

\begin{proof}[{\bf Proof of Claim~\ref{cl:boundbalanced}}]
Observe that 
\begin{eqnarray*}
D_{\cB_f \setminus X_i, V(G)} &=&  \size{\left\{ (u,v)\in E(G): u \in \cB_f\setminus X_i~\mbox{and} \ v \in V(G), \mbox{{\sc Spl}}_i^f(u)=\mbox{{\sc Spl}}_i^f(v)\right\}} \\ &\leq& \size{\{ (u,v) \in E(G): u \in \cB_f \setminus X_i~\mbox{and} \ v \in V(G)\}} 
= \sum_{v \in \cB_f\setminus X_i} \size{N(v)}
%&\leq & \comments{(2 + \Oh(\eps)) \left(\sum_{v \in f^{-1}(L) \cap \cB_f \setminus X_i} \size{N(v)\cap f^{-1}(L)} +  \sum_{v \in f^{-1}(R) \cap \cB_f \setminus X_i } \size{N(v)\cap f^{-1}(R) }\right) +  \frac{\eps n ^2}{10}}
\end{eqnarray*}

As $\cB_f=\cB_f^1 \cup \cB_f^2$, 
\begin{equation}\label{eqn:balancedsum}
D_{\cB_f \setminus X_i, V(G)} \leq  \sum_{v \in \cB_f^1 \setminus X_i} \size{N(v)} + \sum_{v \in \cB_f^2 \setminus X_i} \size{N(v)}.
\end{equation}

We will bound $D_{\cB_f \setminus X_i, V(G)}$ by bounding $\sum\limits_{v \in \cB_f^1 \setminus X_i} \size{N(v)} $ and $\sum\limits_{v \in \cB_f^2 \setminus X_i} \size{N(v)}$ separately, which we prove in the following claim:

\begin{cl}\label{cl:balancedbound}
Let us consider $T_1$ and $T_2$ as follows: $$T_1=  2\left(\sum_{v \in f^{-1}(L) \cap (\cB_f^1 \setminus X_i)}\size{N(v)\cap f^{-1}(L)}+ \sum_{v \in f^{-1}(R) \cap (\cB_f^1 \setminus X_i)}\size{N(v)\cap f^{-1}(R)}\right)+\frac{k\eps n^2}{150},$$
$$ T_2= \left(2 + \frac{k}{200}\right)\left(\sum_{v \in f^{-1}(L) \cap (\cB_f^2 \setminus X_i)} \size{N(v)\cap f^{-1}(L)} +  \sum_{v \in f^{-1}(R) \cap (\cB_f^2 \setminus X_i)} \size{N(v)\cap f^{-1}(R)}\right).$$
Then 
\begin{description}
    \item[(i)] For balanced vertices of {\bf Type 1}, $\sum\limits_{v \in \cB_f^1 \setminus X_i} \size{N(v)} \leq T_1$;
    
    \item[(ii)] For balanced vertices of {\bf Type 2},
$\sum_{v \in \cB_f^2 \setminus X_i} \size{N(v)} \leq T_2.$
\end{description}

\end{cl}

The proof of the above claim is presented in Appendix~\ref{cl:balancedvertices_app}.
Using Claim~\ref{cl:balancedbound} and Equation~\eqref{eqn:balancedsum}, we have the following: 
%From the bounds of $\sum\limits_{v \in \cB_f^1} \size{N(v)}$ and $\sum\limits_{v \in \cB_f^2} \size{N(v)}$, we have the following:
\begin{align*}
&D_{\cB_f \setminus X_i, V(G)} \\
&=  \sum_{v \in \cB_f^1 \setminus X_i} \size{N(v)} + \sum_{v \in \cB_f^2 \setminus X_i } \size{N(v)}  \\
&\leq T_1 + T_2\\
&\leq 2\left(1 + \frac{k}{400}\right)\Pi + \frac{k \eps n^2}{150}.  ~~~~~~~~~~~~~~~~~ (\mbox{From the definitions of $T_1$, $T_2$ and $\Pi$.})
\end{align*}

\remove{
Note that left hand side of Equation~\ref{eqn:balub} represents the crossing edges involving the low vertices. The right hand side gives a lower bound of the bipartite distance of the graph. Thus we have the following:
$$\mbox{number of crossing edges} \leq 3 d_{bip}(G) \leq 3 \eps n^2$$

Similarly, we know that 
$$\sum_{v \in f^{-1}(L) \cap \cB_f^2} \size{N(v)\cap f^{-1}(L)} + \sum_{v \in f^{-1}(R) \cap \cB_f^2} \size{N(v)\cap f^{-1}(R)} \leq \eps n^2$$

So, the total distance with balanced vertices $\leq 4 \eps n^2$.
}
\end{proof}

\subsection*{Proof of Heavy vertex lemma (Lemma~\ref{cl:heavyvertex})}

Before proceeding to prove the Heavy vertex lemma, we will first prove two intermediate claims that will be crucially used in the proof of the lemma. %(Claim ~\ref{cl:largeapprox} and Claim~\ref{cl:smallapprox}).
The first claim states that when we consider a bipartition $f$ of $G$, if a vertex $v \in G$ has a \emph{large} number of neighbors on one side of the partition defined by $f$, the proportion of its neighbors in $X_i$ on the same side of $f$ will be approximately preserved, where $X_i$ is a set of vertices picked at random in {\bf Step-1(i)} of the algorithm \tolbip. The result is formally stated as follows:

\begin{cl}\label{cl:largeapprox}
 Let $f$ be a bipartition of $G$. Consider a vertex $v\in V$.
\begin{description}
\item[(i)]Suppose $\size{N(v)\cap f^{-1}(L)} \geq \frac{k \eps n}{150}$. Then 
$$\mbox{$\size{N(v)\cap f^{-1}(L) \cap X_i} = \left(1 \pm \frac{k}{500}\right) \size{N(v)\cap f^{-1}(L)} \frac{\size{X_i}}{n}$, with probability at least  $1 - o(k\eps)$.}$$

\item[(ii)] Suppose $\size{N(v)\cap f^{-1}(R)} \geq \frac{k \eps n}{150}$. Then $$\mbox{$\size{N(v)\cap f^{-1}(R) \cap X_i} = \left(1 \pm \frac{k}{500}\right) \size{N(v)\cap f^{-1}(R)} \frac{\size{X_i}}{n}$, with probability at least $1-o(k \eps)$.}$$

\end{description}

\end{cl}

The next claim is in similar spirit as that of Claim~\ref{cl:largeapprox}. Instead of considering vertices with large number of neighbors, it considers the case when a vertex has \emph{small} number of neighbors on one side of a bipartition $f$. 

%with respect a bipartition $f$.

\begin{cl}\label{cl:smallapprox}
Let $f$ be a bipartition of $G$. Consider a vertex $v\in V$.
\begin{description}
\item[(i)] Suppose $\size{N(v)\cap f^{-1}(L)} \leq \frac{1}{1 + \frac{k}{200}}\frac{k \eps n}{150}$. Then $$\mbox{$\size{N(v)\cap f^{-1}(L) \cap X_i} \leq \frac{1}{1 + \frac{k}{300}}\frac{k \eps \size{X_i}}{150}$, with probability at least $1-o(k\eps)$.}$$

\item[(ii)] Suppose $\size{N(v)\cap f^{-1}(R)} \leq \frac{1}{1 + \frac{k}{200}}\frac{k \eps n}{150}$. Then $$\mbox{$\size{N(v)\cap f^{-1}(R) \cap X_i} \leq \frac{1}{1 + \frac{k}{300}}\frac{k \eps \size{X_i}}{150}$, with probability at least $1-o(k\eps)$.}$$

\end{description}

\end{cl} 
 
%Claims~\ref{cl:largeapprox} and~\ref{cl:smallapprox} can be proved by using the large deviation inequalities (stated in Appendix~\ref{sec:prob}). For completeness, we give their proofs in Appendix~\ref{sec:completenessrempf_app}.

Claim~\ref{cl:largeapprox} and Claim~\ref{cl:smallapprox} can be proved by using large deviation inequalities (stated in Appendix~\ref{sec:prob}), and the proofs are presented in Appendix~\ref{sec:completenessrempf_app}.

Assuming Claim~\ref{cl:largeapprox} and Claim~\ref{cl:smallapprox} hold, we now prove the Heavy vertex lemma (Lemma~\ref{cl:heavyvertex}).

\begin{proof}[Proof of Lemma~\ref{cl:heavyvertex}]
We will only prove $(i)$ here, which concerns the $L$-heavy vertices. $(ii)$ can be proved in similar fashion.

%Recall that  the Heavy vertex Lemma (Lemma~\ref{cl:heavyvertex}) has two parts: $(i)$ talks about $L$-heavy vertices, and $(ii)$ talks about $R$-heavy vertices. We only present here the proof of $(i)$, $(ii)$ can be proven in similar fashion.

We first characterize $L$-heavy vertices into two categories:

\begin{description}
    \item[(a)] Both $\size{N(v)\cap f^{-1}(L)}$ and $\size{N(v)\cap f^{-1}(R)}$ are large, that is,
    $\size{N(v)\cap f^{-1}(L)} \geq \frac{k\eps n}{150}$ and $\size{N(v)\cap f^{-1}(R)} \geq \frac{1}{1+ \frac{k}{200}}\frac{k\eps n}{150}$. Also, $\size{N(v)\cap f^{-1}(L)} \geq  \left({1+\frac{k}{200}}\right) \size{N(v)\cap f^{-1}(R)}$.
    
     %$$\size{N(v)\cap f^{-1}(L)} \geq \frac{k\eps n}{150} \ \mbox{and} \ \size{N(v)\cap f^{-1}(R)} \geq \frac{1}{1+ \frac{k}{200}}\frac{k\eps n}{150} \ \mbox{and} \ \size{N(v)\cap f^{-1}(L)} \geq \left({1+\frac{k}{200}}\right) \size{N(v)\cap f^{-1}(R)}$$

    \item[(b)]$\size{N(v)\cap f^{-1}(L)}$ is large and $\size{N(v)\cap f^{-1}(R)}$ is small, that is, $\size{N(v)\cap f^{-1}(L)} \geq \frac{k\eps n}{150}$ and $\size{N(v)\cap f^{-1}(R)} {\leq} \frac{1}{1 + \frac{k}{200}}\frac{k\eps n}{150}$.
\end{description}

\begin{description}
\item[Case (a):] %\vspace{\baselineskip}
 Here $\size{N(v)\cap f^{-1}(L)} \geq \left({1+\frac{k}{200}}\right) \frac{k\eps n}{150}$, and $\size{N(v)\cap f^{-1}(R)} \geq \frac{k\eps n}{150}$. From Claim~\ref{cl:largeapprox}, the following hold, with probability at least $1-o(k\eps)$:
 $$ \size{N(v)\cap f^{-1}(L) \cap X_i}=\left(1\pm \frac{k}{500}\right) \size{N(v)\cap f^{-1}(L)} \frac{\size{X_i}}{n}.$$
$$\mbox{and}~ \size{N(v)\cap f^{-1}(R) \cap X_i}=\left(1\pm \frac{k}{500}\right) \size{N(v)\cap f^{-1}(R)} \frac{\size{X_i}}{n}.$$

So, with probability at least $1 -o(k\eps)$, we have the following: 
\begin{eqnarray*}
&&\size{N(v)\cap f^{-1}(L) \cap X_i} - \size{N(v)\cap f^{-1}(R) \cap X_i} \\
&&\geq \left(1-\frac{k}{500}\right)\size{N(v)\cap f^{-1}(L)}\frac{\size{X_i}}{n} - \left(1 + \frac{k}{500}\right)\size{N(v)\cap f^{-1}(R)} \frac{\size{X_i}}{n}\\
 %&\geq& (1-\frac{k}{500})\size{N(v)\cap f^{-1}(L)}\frac{\size{X_i}}{n} - (1 + \frac{k}{500})\left(\size{N(v)\cap f^{-1}(L)} \frac{\size{X_i}}{n} - \frac{k \eps |X_i| }{150}\right)\\
 %&=&-\frac{2k}{500}\size{N(v)\cap f^{-1}(L)} \frac{X_i}{n} + (1+k/500)\frac{k \eps |X_i|}{150} \\
% & \geq& (1+k/500)\frac{k \eps n}{150} - \frac{k}{250} \frac{(1+k/500)\eps X_i}{150}\\
 %&=& (1+k/500) \frac{k}{375}.\\
 &&\geq  \left(1-\frac{k}{500}-\frac{1+\frac{k}{500}}{1+\frac{k}{200}}\right)\frac{    \size{N(v)\cap f^{-1}(L)}\size{X_i}}{ n} \\ 
 &&\quad\quad\quad\quad\quad\quad
 \quad\quad
 \left(\because \size{N(v)\cap f^{-1}(L) } \geq (1 + \frac{k}{200}) \size{N(v)\cap f^{-1}(R)}\right)
\\
&&\geq \frac{k}{1500} \times \frac{k\eps \size{X_i}}{150}\\
&&\geq \frac{k^2\eps \size{X_i}}{225000}
\quad\quad\quad\quad\quad\quad
\quad\quad\quad\quad\quad\quad
\quad\quad\quad
(\because~ k \leq 100)
\end{eqnarray*}

\item[Case (b):]

Here $\size{N(v)\cap f^{-1}(L)} \geq \frac{k\eps n}{150}$  and $\size{N(v)\cap f^{-1}(R)} {\leq} \left(\frac{1}{1+ \frac{k}{200}}\right)\frac{k\eps n}{150}$. So, from Claim~\ref{cl:largeapprox} and Claim~\ref{cl:smallapprox}, the following hold, with probability at least $1 - o(k\eps)$:
$$\size{N(v)\cap f^{-1}(L) \cap X_i}=\left(1\pm \frac{k}{500}\right) \size{N(v)\cap f^{-1}(L)} \frac{\size{X_i}}{n}$$
and
$$
    \size{N(v)\cap f^{-1}(R) \cap X_i} \leq \frac{1}{1 + \frac{k}{300}}\frac{k \eps \size{X_i}}{150}.
$$

Thus, with probability at least $1 -o(k\eps)$, we have the following:
\begin{align*}
&\size{N(v)\cap f^{-1}(L) \cap X_i} - \size{N(v)\cap f^{-1}(R) \cap X_i} &\\
&\geq (1-\frac{k}{500})\size{N(v)\cap f^{-1}(L)}\frac{\size{X_i}}{n} -\frac{1}{1 + \frac{k}{300}} \frac{k \eps \size{X_i}}{150} &\\ 
&=(1-\frac{k}{500}) \frac{k \eps \size{X_i}}{150}- \frac{1}{1 + \frac{k}{300}} \frac{k \eps \size{X_i}}{150} &\\
&\geq \frac{1}{1500}\left(2k-\frac{k^2}{100}\right)\frac{k\eps \size{X_i}}{150} &\\
&\geq \frac{k^2\eps \size{X_i}}{225000} & (\because~ k \leq 100)
\end{align*}
\end{description}

This completes the proof of part $(i)$ of Lemma~\ref{cl:heavyvertex}.

\end{proof}

\subsection{Proof of Soundness}\label{sec:correctness2}

%Z as random pair of vertices.

%========

%The soundness proof is fairly standard and relies on the "for all"-condition for being $\epsilon$-far from bipartite. \textcolor{green}{pls verify}
In this section, we prove the following theorem:
\begin{theo}\label{theo:soundness}
Let us assume that $G$ is $(2+k) \eps$-far from being bipartite. Then 
\tolbip reports the same, with probability at least $\frac{9}{10}$.
\end{theo}

%Let us now consider the case when $G$ is $10 \eps$-far from being bipartite. Now we would like to prove that our algorithm reports the same with probability at least $\frac{2}{3}$.

Assume $f$ be a bipartition of $V(G)$. Now let us consider a {\sc derived} bipartition $\mbox{{\sc Der}}_i^f$ with respect to $f$ by $f_{ij}$, and choose a set of random vertex pairs $Y$ such that $|Y|=|Z|$. 
Let $\chi_{ij}^f$ denote the fraction of vertex pairs of $Y$ that are monochromatic with respect to the bipartition $\mbox{{\sc Der}}_i^f$, that is,
$$\chi_{ij}^f = 2 \cdot  \frac{\size{\left\{\{a,b\}\in Y: \{a,b\}\in E(G)~\mbox{and}~\mbox{{\sc Der}}_i^f(a)=\mbox{{\sc Der}}_i^f(b)\right\}}}{\size{Y}}.$$
%%%%%%%%%

%%%%%%%%%%%%%%
\begin{obs}\label{obs:chiub}
$\chi^f_{ij} \leq \left(2+\frac{k}{20}\right) \eps$ holds with probability at most $\frac{1}{10N}$, where $N= 2^{\Oh(\frac{1}{k^3\eps} \log \frac{1}{k \eps})}$.
\end{obs}
\begin{proof}
Since $G$ is $(2+k) \eps$-far from being bipartite, the same holds for the bipartition $\mbox{{\sc Der}}_i^f$ as well, that is, $d_{bip}(G,\mbox{{\sc Der}}_i^f) \geq (2+k) \eps n^2.$ So, $\E[\chi^f_{ij}]\geq \left(2+k\right) \eps$. Using Chernoff bound (see Lemma~\ref{lem:cher_bound3}), we can say that,
$\pr \left(\chi^f_{ij} \leq \left( 2+ \frac{k}{20}\right) \eps \right) \leq  \frac{1}{10 N}$. Since $\size{Z}=\Oh\left(\frac{1}{k^5 \eps^2 }\log {\frac{1}{k \eps}}\right)$, the result follows.
\end{proof}

We will be done with the proof by proving the following claim, that says that bounding $\chi^f _{ij} $ is equivalent to bounding $\zeta_{ij}$.

\begin{cl}\label{cl:zetachisoundness}
For any $i \in [t]$, and any $f_{ij}\in \cF_i$, the probability distribution of $\zeta_{ij}$ is identical to that of $\chi^f_{ij}$ for some {\sc derived} bipartition with respect to $f$ by $f_{ij}$.

%with a probability of at least $1/2$.

%For any $i \in [t]$, there exists $f_{ij}\in \cF_i$ such that the probability distribution of $\zeta_{ij}$ is identical to that of $\chi_{ij}$

\end{cl}
\begin{proof}
Consider a bipartition $f_{ij} \in \cF_i$ of $X_i$, and the bipartition $f^{'}_{ij}$ of $X_i\cup Z$, constructed by extending $f_{ij}$, as described in the algorithm. For the sake of the argument, let us construct a new bipartition $f_{ij}^{''}$ of $V(G)$ by extending the bipartition $f'_{ij}$, following the same rule of {\bf Step-2~(ii)~(b)} of the algorithm. Observe that $f_{ij}^{''}(v)=f_{ij}(v)$, for each $v \in X_i$. Thus $f_{ij}^{''}$ is a {\sc derived} bipartition with respect to some $f$ by $f_{ij}$.
Hence, the claim follows according to the way we generate $\zeta_{ij}$, along with the fact that $Z$ is chosen uniformly at random by the algorithm in {\bf Step-1~(ii)}.

% For the sake of argument, let us extend the domain of $f_{ij}^{'}$ to $V(G)$, by the rule (in {\bf Step-2(ii)(b)}) that is used to map the vertices of $X_i \cup Z$ for constructing $f_{ij}'$. We refer to this bipartition of $V(G)$ as $f_{ij}^{''}$.

% Observe that $f_{ij}^{''}(v)=f_{ij}(v)$ for each $v \in X_i$. So, $f_{ij}^{''}$ is a {\sc derived} bipartition with respect to some $f$ by $f_{ij}$. %{Since $G$ is $16\eps$-far, bipartite distance with respect to $f_{ij}''$ must also be at least $16\epsilon$}.

% \remove{Recall the procedure of determining $\zeta_{ij}$ as described in {\bf Step 2} of our algorithm presented in Section~\ref{sec:algosec}. 
%   Let us think of the following instance (for argument purpose only). Let $f_{ij}^{''}$ is a bipartition from $V(G)$, where for each $v \in V(G)$, $f_{ij}^{''}(v) $ is determined by the same rule that we are using to determine $f_{ij}^{''}(z)$ for each $z \in Z$.}
\end{proof}

%The proof of the above claim follows since when $G$ is $10 \eps$-far from being bipartite, it holds for all possible bipartitions, thus for any derived bipartition $f$ by $f_{ij}$ as well.

%\textcolor{blue}{Sayantan: Need to write something about random restriction.}

\remove{
\color{blue}
As $t =\Oh(\log \frac{1}{\eps})$, the above claim implies that there exists an $i \in [t]$ and $f_{ij}\in \cF_i$ such that the probability distribution of $\zeta_{ij}$ is identical to that of $\chi_{ij}$, with a probability of at least $1-o(\eps)$.

The proof of Claim~\ref{cl:zetachisoundness} follows in similar line to that of Claim~\ref{cl:zetachi} by considering derived bipartitions instead of special bipartitions. We omit the proof here.
}
\color{black}

Let us now define a pair $(X_i, f_{ij})$, with $i \in [t]$ and $f_{ij} \in \mathcal{F}_i$ as a {\bf configuration}. Now we make the following observation which follows directly from the description of the algorithm.

\begin{obs}\label{obs:config}
Total number of possible configurations is $N= 2^{\Oh(\frac{1}{k^3\eps} \log \frac{1}{k \eps})}$.
\end{obs}

%\textcolor{blue}{do we need to add a line for the proof?}

Note that Claim~\ref{cl:zetachisoundness} holds for a particular $f_{ij} \in \cF_i$. Recall that in {\bf Step-2(iii)}, our algorithm \tolbip reports that $G$ is $(2+k) \eps$-far if $\zeta_{ij} > \left(2+\frac{k}{20}\right) \eps$, for all $i \in [t]$ and $f_{ij} \in \cF_i$. So, using the union bound, along with Observation~\ref{obs:chiub}, Claim~\ref{cl:zetachisoundness} and Observation~\ref{obs:config}, we are done with the proof of Theorem~\ref{theo:soundness}.

\section{Conclusion}\label{sec:conclusion}
%\paragraph{The bottleneck of our technique to improve the bound in Theorem~\ref{theo:mainoverview}}
We believe that our result will certainly improve the current understanding of (tolerant) bipartite testing in the dense graph model. However, 
one may wonder whether the analysis can be improved to show that the algorithm (presented in Section~\ref{sec:algo_formal}) can decide whether $d_{bip}(G) \leq \eps n^2$ or $d_{bip}(G) \geq c \eps n^2$ for any $c>1$. There is a bottleneck in our technique as we are bounding error due to the balanced vertices by the sum of degrees of the balanced vertices (as done in Claim~\ref{cl:boundbalanced}). 
Because of this reason, it is not obvious if our algorithm (and its analysis) 
can be used to get a result, like of Theorem~\ref{theo:mainoverview_algo}, for all $c>1$  
with the same query complexity.

On a different note, we can decide $d_{bip}(G) \leq \eps n^2$ or $d_{bip}(G) \geq (1+k) \eps n^2$ by using $\tOh\left(\frac{1}{k^6 \eps^6}\right)$ queries, which can be derived from the work of Alon, Vega, Kannan and Karpinski~\cite{DBLP:journals/jcss/AlonVKK03} (see Corollary~\ref{coro:two} in Appendix~\ref{sec:alon}). Hence, any algorithm that solves the general bipartite distance problem 
with query complexity $o\left(\frac{1}{k^6 \eps^6}\right)$, will be of huge interest.

\paragraph*{Acknowledgement.}
The authors would like to thank Yufei Zhao, Dingding Dong and Nitya Mani for pointing out a mistake in an earlier version of this paper. 
Gopinath Mishra's research is supported in part by the Centre for Discrete Mathematics and its Applications (DIMAP) and by EPSRC award EP/V01305X/1.
%as well as the reviewers of ICALP for various suggestions that improved the presentation of the paper.

%\newpage
\bibliographystyle{alpha}
\bibliography{reference}
\appendix

\newpage
\section{Remaining Proofs of Section~\ref{sec:correctness}}\label{sec:completenessrempf_app}
%\color{blue}
Here we include proofs of four claims that were not formally proven in Section~\ref{sec:correctness}.

\begin{cl}[Restatement of Claim~\ref{cl:balancedbound} (i)]\label{cl:balancedvertices_app}
Let $$T_1=  2\left(\sum_{v \in f^{-1}(L) \cap (\cB_f^1 \setminus X_i)}\size{N(v)\cap f^{-1}(L)}+ \sum_{v \in f^{-1}(R) \cap (\cB_f^1 \setminus X_i)}\size{N(v)\cap f^{-1}(R)}\right)+\frac{k\eps n^2}{150}.$$ Then,  for balanced vertices of {\bf Type 1}, $\sum\limits_{v \in \cB_f^1 \setminus X_i} \size{N(v)} \leq T_1$.
\end{cl}

\begin{proof}

%\paragraph*{Bounding $\sum\limits_{v \in \cB_f^1 \setminus X_i} \size{N(v)} $:}
Let us consider an optimal bipartition $f$. Then, for any vertex $v \in f^{-1}(L) \cap (\cB_f^1 \setminus X_i)$, we can say the following:
$$\frac{-k\eps n}{150} \leq \size{N(v)\cap f^{-1}(L)} - \size{N(v)\cap f^{-1}(R)} \leq 0$$

Thus
$$\frac{-k\eps n \size{f^{-1}(L) \cap (\cB_f^1 \setminus X_i)}}{150} \leq \sum_{v \in f^{-1}(L) \cap (\cB_f^1 \setminus X_i)}\size{N(v)\cap f^{-1}(L)} - \sum_{v \in f^{-1}(L) \cap (\cB_f^1 \setminus X_i)} \size{N(v)\cap f^{-1}(R)} \leq 0 $$

Similarly, we can also say that,
$$\frac{-k \eps n \size{f^{-1}(R) \cap (\cB_f^1 \setminus X_i)}}{150} \leq \sum_{v \in f^{-1}(R) \cap (\cB_f^1 \setminus X_i)}\size{N(v)\cap f^{-1}(R)} - \sum_{v \in f^{-1}(R) \cap (\cB_f^1 \setminus X_i)} \size{N(v)\cap f^{-1}(L)} \leq 0.$$

Since $f^{-1}(L) \cup f^{-1}(R) = V(G)$, and $ f^{-1}(L) \cap f^{-1}(R)= \emptyset$, we have the following {four inequalities:}
\begin{eqnarray*}
&&\frac{-k\eps n \size{\cB_f^1 \setminus X_i}}{150} \; \leq \; \left(\sum_{v \in f^{-1}(L) \cap (\cB_f^1 \setminus X_i)}\size{N(v)\cap f^{-1}(L)}+ \sum_{v \in f^{-1}(R) \cap (\cB_f^1 \setminus X_i)}\size{N(v)\cap f^{-1}(R)}\right) \nonumber\\
&&~~~~~~~~~~~~~~~~~~~~~~~~-\left(\sum_{v \in f^{-1}(L) \cap (\cB_f^1 \setminus X_i)}  \size{N(v)\cap f^{-1}(R)}+\sum_{v \in f^{-1}(R) \cap (\cB_f^1 \setminus X_i)} \size{N(v)\cap f^{-1}(L)}\right)
\end{eqnarray*}

\begin{eqnarray*}
\mbox{So,} &&\sum_{v \in f^{-1}(L) \cap (\cB_f^1 \setminus X_i)} \size{N(v)\cap f^{-1}(R)}+\sum_{v \in f^{-1}(R) \cap (\cB_f^1 \setminus X_i)} \size{N(v)\cap f^{-1}(L)}  \nonumber\\
&\leq&~~~\sum_{v \in f^{-1}(L) \cap (\cB_f^1 \setminus X_i)}\size{N(v)\cap f^{-1}(L)}+ \sum_{v \in f^{-1}(R) \cap (\cB_f^1 \setminus X_i)}\size{N(v)\cap f^{-1}(R)}+\frac{k \eps n \size{\cB_f^1 \setminus X_i}}{150}
\end{eqnarray*}

\begin{eqnarray*}
&&\mbox{Thus,}\sum_{v \in f^{-1}(L) \cap (\cB_f^1 \setminus X_i)} \size{N(v)}+\sum_{v \in f^{-1}(R) \cap (\cB_f^1 \setminus X_i)} \size{N(v)}  \nonumber\\
&&~~~~~~~\; \leq \; 2\left(\sum_{v \in f^{-1}(L) \cap (\cB_f^1 \setminus X_i)}\size{N(v)\cap f^{-1}(L)}+ \sum_{v \in f^{-1}(R) \cap (\cB_f^1 \setminus X_i)}\size{N(v)\cap f^{-1}(R)}\right)+\frac{k \eps n^2}{150}
\end{eqnarray*}

So, we have the following:
\begin{eqnarray*}
\sum_{v \in  \cB_f^1 \setminus X_i}\size{N(v)} \; \leq \; T_1.
\end{eqnarray*}
\end{proof}

\begin{cl}[Restatement of Claim~\ref{cl:balancedbound}(ii)]

Let 
$$ T_2= \left(2 + \frac{k}{200}\right)\left(\sum_{v \in f^{-1}(L) \cap (\cB_f^2 \setminus X_i)} \size{N(v)\cap f^{-1}(L)} +  \sum_{v \in f^{-1}(R) \cap (\cB_f^2 \setminus X_i)} \size{N(v)\cap f^{-1}(R)}\right).$$
Then, for balanced vertices of {\bf Type 2},
$\sum\limits_{v \in \cB_f^2 \setminus X_i} \size{N(v)} \leq T_2.$
\end{cl}

\begin{proof}

%\paragraph*{Bounding $\sum\limits_{v \in \cB_f^2 \setminus X_i} \size{N(v)} $:}

Recall the definition of balanced vertices of {\bf Type 2} from Definition~\ref{defi:balancedvertex}. Summing over all the vertices of $f^{-1}(L) \cap (\cB_f^2 \setminus X_i)$, we have
\begin{eqnarray*}
\sum_{v \in f^{-1}(L) \cap (\cB_f^2 \setminus X_i)} \size{N(v)\cap f^{-1}(L)} &\leq& \sum_{v \in f^{-1}(L) \cap (\cB_f^2 \setminus X_i)} \size{N(v)\cap f^{-1}(R)}\\ &\leq& 
\left(1+\frac{k}{200}\right) \sum_{v \in f^{-1}(L) \cap (\cB_f^2 \setminus X_i)} \size{N(v)\cap f^{-1}(L)}
\end{eqnarray*}

Similarly, we can also say that
\begin{eqnarray*}
\sum_{v \in f^{-1}(R) \cap (\cB_f^2 \setminus X_i)} \size{N(v)\cap f^{-1}(R)} &\leq& \sum_{v \in f^{-1}(R) \cap (\cB_f^2 \setminus X_i)} \size{N(v)\cap f^{-1}(L)}\\ &\leq& \left(1+\frac{k}{200}\right) \sum_{v \in f^{-1}(R) \cap (\cB_f^2 \setminus X_i)} \size{N(v)\cap f^{-1}(R)}.
\end{eqnarray*}

Summing the above two inequalities, we get the following {three inequalities:}
\begin{eqnarray*}\label{eqn:balub}
&&\sum_{v \in f^{-1}(L) \cap (\cB_f^2 \setminus X_i)} \size{N(v)\cap f^{-1}(R)} + \sum_{v \in f^{-1}(R) \cap (\cB_f^2 \setminus X_i)} \size{N(v)\cap f^{-1}(L)}\nonumber\\ 
&&\;\leq\; \left(1 + \frac{k}{200}\right) \left(\sum_{v \in f^{-1}(L) \cap (\cB_f^2 \setminus X_i)} \size{N(v)\cap f^{-1}(L)} +  \sum_{v \in f^{-1}(R) \cap (\cB_f^2 \setminus X_i)} \size{N(v)\cap f^{-1}(R)}\right) 
\end{eqnarray*}

\begin{eqnarray*}
&&\mbox{So,}\sum_{v \in f^{-1}(L) \cap (\cB_f^2 \setminus X_i)} \size{N(v)}+\sum_{v \in f^{-1}(R) \cap (\cB_f^2 \setminus X_i)} \size{N(v)} \nonumber\\
&&\leq \left(2 + \frac{k}{200}\right)  \left(\sum_{v \in f^{-1}(L) \cap (\cB_f^2 \setminus X_i)} \size{N(v)\cap f^{-1}(L)} +  \sum_{v \in f^{-1}(R) \cap (\cB_f^2 \setminus X_i)} \size{N(v)\cap f^{-1}(R)}\right) .
\end{eqnarray*}

Thus, we have the following:
\begin{eqnarray*}
\sum_{v \in (\cB_f^2 \setminus X_i)} \size{N(v)} \; \; \leq \; \; T_2 .
\end{eqnarray*}

% \remove{
% \color{blue}
% \begin{eqnarray*}\label{eqn:balub}
% &&\sum_{v \in f^{-1}(L) \cap \cB_f} \size{N(v)\cap f^{-1}(R)} + \sum_{v \in f^{-1}(R) \cap \cB_f} \size{N(v)\cap f^{-1}(L)}\nonumber\\ 
% &&~~~~~~~~~~~~~~~~~\;\leq\; (1 + \Oh(\eps)) \left(\sum_{v \in f^{-1}(L) \cap \cB_f} \size{N(v)\cap f^{-1}(L)} +  \sum_{v \in f^{-1}(R) \cap \cB_f} \size{N(v)\cap f^{-1}(R)}\right) +  \frac{\eps n \size{\cB_f^1 \setminus X_i}}{10}  
% \end{eqnarray*}

% \begin{eqnarray}
% \sum_{v \in \cB_f \setminus X_i} \size{N(v)} \; \; \leq \; \; (2 + \Oh(\eps)) \left(\sum_{v \in f^{-1}(L) \cap \cB_f \setminus X_i} \size{N(v)\cap f^{-1}(L)} +  \sum_{v \in f^{-1}(R) \cap \cB_f \setminus X_i } \size{N(v)\cap f^{-1}(R) }\right) +  \frac{\eps n \size{\cB_f^1 \setminus X_i}}{10} 
% \end{eqnarray} }
\color{black}

\end{proof}

\color{black}
\begin{cl}[Restatement of Claim~\ref{cl:largeapprox}]\label{cl:largeapprox_app}
 Let $f$ be a bipartition of $G$. Consider a vertex $v\in V$.
\begin{description}
\item[(i)]Suppose $\size{N(v)\cap f^{-1}(L)} \geq \frac{k \eps n}{150}$. Then $\size{N(v)\cap f^{-1}(L) \cap X_i} = \left(1 \pm \frac{k}{500}\right) \size{N(v)\cap f^{-1}(L)} \frac{\size{X_i}}{n}$ holds, with probability at least $1 - o(k\eps)$.

\item[(ii)] Suppose $\size{N(v)\cap f^{-1}(R)} \geq \frac{k \eps n}{150}$. Then $\size{N(v)\cap f^{-1}(R) \cap X_i} = \left(1 \pm \frac{k}{500}\right) \size{N(v)\cap f^{-1}(R)} \frac{\size{X_i}}{n}$ holds, with probability at least $1-o(k \eps)$.

\end{description}

\end{cl}

\begin{proof}
We prove only part $(i)$ of the claim. Part $(ii)$ can be proven analogously.

From the condition stated in $(i)$, we know that
$$\size{N(v)\cap f^{-1}(L)} \geq \frac{k \eps n}{150}.$$

Since $X_i$ is chosen randomly, we can say that
$$\E\left[\size{N(v)\cap f^{-1}(L) \cap X_i}\right] \geq \frac{k \eps \size{X_i}}{150}.$$

Using Chernoff bound (see Lemma \ref{lem:cher_bound3}), we have
$$\pr\left(\size{N(v)\cap f^{-1}(L) \cap X_i} \neq (1 \pm \frac{k}{500}) \size{N(v)\cap f^{-1}(L)} \frac{\size{X_i}}{n} \right) \leq 2e^{-\Omega\left( k^3\eps \size{X_i}\right)} =  o(k\eps)$$

The last inequality follows from the fact that $\size{X_i} =\Oh( \frac{1}{k^3\eps}\log \frac{1}{k\eps})$.

%Similarly, we can also prove that
%$$\pr\left(\size{N(v)\cap f^{-1}(L) \cap X_i} \geq (1+ \frac{1}{50}) \size{N(v)\cap f^{-1}(L)} \frac{\size{X_i}}{n} \right) \leq e^{\frac{- \eps \size{X_i}}{7500}} \leq o(1)$$

%$$\pr\left(\size{N(v)\cap f^{-1}(L) \cap X} \leq E[\size{N(v)\cap f^{-1}(L) \cap X}] - \frac{\eps \size{X}}{20} \right) \leq e^{\frac{- \eps \size{X}}{40}} \leq \eps^c$$

%\begin{eqnarray*}
%&&\size{N(v)\cap f^{-1}(L)} \geq \frac{\eps n}{10} \\ &&\E[\size{N(v)\cap f^{-1}(L) \cap X}] \geq \frac{\eps \size{X}}{10} \\ &&\pr(\size{N(v)\cap f^{-1}(L) \cap X} \leq E[\size{N(v)\cap f^{-1}(L) \cap X}] - \frac{\eps \size{X}}{20} ) \leq e^{\frac{- \eps \size{X}}{40}}
%\end{eqnarray*}

\end{proof}

\begin{cl}[Restatement of Claim~\ref{cl:smallapprox}]\label{cl:smallapprox_app}
Let $f$ be a bipartition of $G$. Consider a vertex $v\in V$.
\begin{description}
\item[(i)] Suppose $\size{N(v)\cap f^{-1}(L)} \leq \frac{1}{1+ \frac{k}{200}}\frac{k\eps n}{150}$. Then $\size{N(v)\cap f^{-1}(L) \cap X_i} \leq \frac{1}{1+\frac{k}{300}}\frac{k \eps \size{X_i}}{150}$ holds, with probability at least $1-o(k\eps)$.

\item[(ii)] Suppose $\size{N(v)\cap f^{-1}(R)} \leq \frac{1}{1+ \frac{k}{200}}\frac{k\eps n}{150}$. Then $\size{N(v)\cap f^{-1}(R) \cap X_i} \leq \frac{1}{1+\frac{k}{300}}\frac{k \eps \size{X_i}}{150}$ holds, with probability at least $1-o(k\eps)$.

\end{description}

\end{cl}

\begin{proof}
%\complain{[To be edited]}
We will only prove part $(i)$ here. Part $(ii)$ can be proven in similar manner.

From the condition stated in $(i)$, we know that $$\size{N(v)\cap f^{-1}(R)} \leq \frac{1}{1+\frac{k}{200}}\frac{(1+k)\eps n}{150}.$$

Since $X_i$ is chosen at random, we can say that
$$\E\left[\size{N(v)\cap f^{-1}(R) \cap X_i}\right] \leq \frac{1}{1+\frac{k}{200}}\frac{(1+k)\eps \size{X_i}}{150}.$$

Using Chernoff bound (see Lemma~\ref{lem:cher_bound3}), we have
$$\pr\left(\size{N(v)\cap f^{-1}(L) \cap X_i} \geq \frac{1}{1+\frac{k}{300}}\frac{(1+k) \eps \size{X_i}}{150} \right) \leq e^{-\Omega\left(k^2\eps \size{X_i}\right)} \leq o(k \eps)$$

The last inequality follows due to the fact that $\size{X_i} =\Oh( \frac{1}{k^3\eps}\log \frac{1}{k \eps})$.
\end{proof}

%\newpage
\section{Algorithm for bipartite distance estimation with query complexity $\widetilde{\Oh}\left(\frac{1}{\varepsilon^6}\right)$}~\label{sec:alon}
Formally, we state the following theorem.
\begin{theo}\label{theo:bipest_app}
Given an unknown graph $G$ on $n$ vertices and any approximation parameter $\eps \in (0,1)$, there is an algorithm that performs $\tOh(\frac{1}{\eps^6})$ adjacency queries, and outputs a number $\widehat{d}_{bip}(G)$ such that, with probability at least $\frac{9}{10}$, the following holds:
$$
    d_{bip}(G) - \eps n^2 \leq \widehat{d}_{bip}(G) \leq d_{bip}(G)+ \eps n^2,
$$
where $d_{bip}(G)$ denotes the bipartite distance of $G$.
%~\footnote{The success probability can be improved to any $1- \delta$ with an extra multiplicative factor of $\log \frac{1}{\delta}$. Also, note that the additive approximation parameter should be $\eps {n \choose 2}$. However, for simplicity of presentation, we will take $\eps n^2$ instead of $\eps {n \choose 2}.$ }.
\end{theo}
We have the following two corollaries of the above theorem.
\begin{coro}\label{coro:one}
There exists an algorithm that given adjacency query access to a graph $G$ with $n$ vertices and a parameter $\eps \in (0,1)$ such that, with probability at least $\frac{9}{10}$, decides whether $d_{bip}(G) \leq \eps n^2$ or $d_{bip}(G) \geq (2+\Omega(1)) \eps n^2$  using $\tOh\left(\frac{1}{\eps^6}\right)$ many queries to the adjacency matrix of $G$. 
\end{coro}
\begin{coro}\label{coro:two}
There exists an algorithm that given adjacency query access to a graph $G$ with $n$ vertices and a parameter $\eps \in (0,1)$ such that, with probability at least $\frac{9}{10}$, decides whether $d_{bip}(G) \leq \eps n^2$ or $d_{bip}(G) \geq (1+k) \eps n^2$  using $\tOh\left(\frac{1}{k^6\eps^6}\right)$ many queries to the adjacency matrix of $G$. 
\end{coro}
To prove Theorem~\ref{theo:bipest_app}, we first discuss the connection between \maxcut and bipartite distance of a graph $G$. Then we use the result for \maxcut estimation by Alon, Vega, Kannan and Karpinski~\cite{DBLP:journals/jcss/AlonVKK03}.

\paragraph*{Connection between \maxcut and $d_{bip}(G)$:}

%The bipartite distance $d_{bip}(G)$ of a graph $G$ is  defined as the minimum number of edges we need to remove to make $G$ bipartite. See Definition~\ref{defi:bipdist} for a more formal definition of bipartite distance. 

For a graph $G = (V, E)$ on the {\em vertex} set $V$ and {\em edge} set $E$, let $S$ be a subset of $V$. We define 
\begin{eqnarray*}
   \mbox{ {\sc Cut}}(S) : = \ \mid \{ \{u,v\} \in E \; \mid\; \size{\{u,v\} \cap S} = 1  \} \mid
\end{eqnarray*}
\emph{Maximum Cut} (henceforth termed as {\sc MaxCut}), denoted by $M(G)$, is a partition of the vertex set $V$ of $G$ into two parts such that the number of edges crossing the partition is maximized, that is, 
$$
    M(G) : = \max_{S \subseteq V} \mbox{{\sc Cut}}(S).
$$

The following equation connects \maxcut and the bipartite distance of a graph $G$:
\begin{equation}\label{main-equation}
    d_{bip}(G)  = \size{E(G)} - M(G).
\end{equation}
So, $d_{bip}(G)$ can be estimated by estimating $\size{E(G)}$ and $M(G)$. 

\paragraph*{Result on edge estimation:}

Observe that estimating $\size{E(G)}$ with $\eps n^{2}$ additive error is equivalent to {\em parameter estimation problem} in probability theory, see Mitzenmacher and Upfal~\cite[Section~4.2.3]{mitzenmacher2017probability}.

\begin{pre}[Folklore]\label{obs:edgeestimation}
Given any graph $G$ on $n$ vertices and an input parameter $\eps \in (0,1)$, the size of the edge set $E(G)$ can be estimated within an additive $\eps n^2$ error, with probability at least $\frac{9}{10}$, using $\Oh(\frac{1}{\eps^2})$ many adjacency queries to $G$. 
%Note that time and sample complexities of the algorithm will also be $\Oh\left( \frac{1}{\eps^{2}}\right)$.}
\end{pre}

\subsection*{\maxcut estimation by using $\tOh\left(\frac{1}{\eps^6}\right)$ queries:}

Let $G= (V, E)$ be an $n$ vertex graph. 
Both Alon, Vega, Kannan and Karpinski~\cite{DBLP:journals/jcss/AlonVKK03} and  Mathieu and Schudy~\cite{DBLP:conf/soda/MathieuS08} showed that if $S$ is a $t$-sized random subset of $V$, where $t = O\left( \frac{1}{\eps^{4}} \log \frac{1}{\eps}\right)$, then, with probability at least $\frac{9}{10}$, we have the following:
$$
    \size{\frac{M(G\mid_{S})}{t^{2}} - \frac{M(G)}{n^{2}}} \leq \frac{\eps}{2}
$$
where $G\mid_{S}$ denotes the induced graph of $G$ on the vertex set $S$. So, the above inequality tells us that if we can get an $\eps t^{2}/2$ additive error to $M(G\mid_{S})$, then we can get an $\eps n^{2}$ additive estimate for $M(G)$. Observation~\ref{obs:maxcutestmination} implies that using $O\left( \frac{t}{\eps^{2}}\right) = O\left( \frac{1}{\eps^{6}}\log \frac{1}{\eps}\right)$ many adjacency queries to $G\mid_{S}$, we can get an $\frac{\eps t^{2}}{2}$ additive estimate to $M(G\mid_{S})$. Therefore, the query complexity of \maxcut algorithms of Alon, Vega, Kannan and Karpinski~\cite{DBLP:journals/jcss/AlonVKK03} and  Mathieu and Schudy~\cite{DBLP:conf/soda/MathieuS08} is at most $O\left( \frac{1}{\eps^{6}} \log \frac{1}{\eps}\right)$.

Now we state and prove the following observation.
\begin{obs}[Folklore]\label{obs:maxcutestmination}
For a graph $G$ with $n$ vertices and an approximation parameter $\eps \in (0,1)$, $\Theta\left( \frac{n}{\eps^{2}} \right)$ many adjacency queries to $G$ are sufficient to get an $\eps n^{2}$ additive approximation to \maxcut $M(G)$, with probability at least $\frac{9}{10}$.
\end{obs}

\begin{proof}

We sample $t$ many pairs of vertices $\{a_{1}, b_{1}\}, \dots, \, \{a_{t}, b_{t}\}$ uniformly at random and independent of each other, where $t = \Theta(\frac{n}{\eps^2})$.  %Let us assume that $(S,T)$ is the \maxcut of $G$ and we have sampled $t$ many pairs of vertices $\{a_{1}, b_{1}\}, \dots, \, \{a_{t}, b_{t}\}$ uniformly at random, independent of each other. 
Thereafter, we perform $t$ many adjacency queries to those sampled pairs of vertices. Now fix a subset $S \subset V(G)$  and let us denote $(S,\overline{S})$ to be the set of edges between $S$ and $\overline{S}$.

Let us now define a set of random variables,  one for each sampled pair of vertices as follows:

\[  X_i= \left\{
\begin{array}{ll}
     
      1, & \mbox{if} \ \{a_i,b_i\} \in (S,\overline{S}) \\
      \vspace{3pt}
            0, & \mbox{Otherwise}  
\end{array} 
\right. \]

We will output $\max\limits_{S \subset V(G)} \widehat{M}_S$ as our estimate of $M(G)$, where $\widehat{M}_S=\frac{{n \choose 2}}{t} \sum\limits_{i=1}^{t} X_i.$

Let us denote $X= \sum\limits_{i=1}^t X_i.$ Note that %$$\E\left[X_i\right]=\pr\left(X_i = 1\right) = \frac{\size{(S \times T) \cap E}}{\size{E}}.$$
$$\E\left[X_i\right]=\pr\left(X_i = 1\right) = \frac{\size{(S,\overline{S})}}{{n \choose 2}}, \mbox{and hence }~  \E\left[\widehat{M}_S\right] = \frac{{n \choose 2}}{t} \E\left[\sum_{i=1}^{t} X_i\right] = \size{(S,\overline{S})} $$

Using Hoeffding's Inequality (See Lemma~\ref{lem:hoeffdingineq}), we can say that

%$$\pr\left(\size{M -\widehat{M}}  \geq \frac{\eps n^2}{10}\right) = \pr\left(\size{ X - \E[X]}  \geq \frac{\eps t}{10}\right)  \leq 2 e^{-\Theta(\frac{\eps^2 t^2}{t})} \leq 2 e^{-\Theta(n)}$$
$$\pr\left(\size{\size{(S,\overline{S})} -\widehat{M}_S}  \geq \frac{\eps n^2}{10}\right) \leq  \pr\left(\size{ X - \E[X]}  \geq \frac{\eps t}{10}\right)  \leq 2 e^{-\Theta(\frac{\eps^2 t^2}{t})} \leq 2 e^{-\Theta(n)}.$$

Using union bound over all $S \subset V(G)$, we can show that with probability at least $3/4$, for each $S \subset V(G)$, $\widehat{M}_S$ approximates $\size{(S, \overline{S})}$ with $\eps n^2$ additive error. Therefore $\max\limits_{S \subset V(G)} \widehat{M}_S$ estimates $M(G)$ with  additive error $\eps n^2$, with probability at least $3/4$.
%The last inequality follows from the fact that $t = \Theta(\frac{n}{\eps^2})$.
%Using the union bound over all possible maxcuts as $(S,T)$, we have the result.
\end{proof}

%\end{rem}

\section{Large Deviation Inequalities}\label{sec:prob}

\remove{
\begin{lem}[Chernoff-Hoeffding bound, see~\cite{dubhashi2009concentration}]
\label{lem:cher_bound1}
Let $X_1, \ldots, X_n$ be independent random variables such that $X_i \in [0,1]$. For $X=\sum\limits_{i=1}^n X_i$ and $\mu=\E[X]$, the following holds for all $0\leq \delta \leq 1$
$$ 
    \pr\left(\size{X-\mu} \geq \delta\mu\right) \leq 2\exp{\left(\frac{-\mu \delta^2}{3}\right)}.
$$

\end{lem}
\begin{lem}[Chernoff-Hoeffding bound, see~\cite{dubhashi2009concentration}]
\label{lem:cher_bound2}
Let $X_1, \ldots, X_n$ be independent random variables such that $X_i \in [0,1]$. For $X=\sum\limits_{i=1}^n X_i$ and $\mu_l \leq \E[X] \leq \mu_h$, the followings hold for any $\delta >0$:
\begin{itemize}
\item[(i)] $\pr \left( X \geq \mu_h + \delta \right) \leq \exp{\left(\frac{-2\delta^2}{n}\right)}$.
\item[(ii)] $\pr \left( X \leq \mu_l - \delta \right) \leq \exp{\left(\frac{-2\delta^2}{n}\right)}$.
\end{itemize}

\end{lem}}

\begin{lem}[Chernoff-Hoeffding bound, see~\cite{dubhashi2009concentration}]
\label{lem:cher_bound3}
Let $X_1, \ldots, X_n$ be independent random variables such that $X_i \in [0,1]$. For $X=\sum\limits_{i=1}^n X_i$ and $\mu_l \leq \E[X] \leq \mu_h$, the followings hold for any $ 0 < \eps < 1$:
\begin{itemize}
\item[(i)] $\pr \left( X \geq (1 + \eps)\mu_h \right) \leq \exp{\left(\frac{-\eps^2 \mu_h}{3}\right)}$.
\item[(ii)] $\pr \left( X \leq (1 - \eps)\mu_l \right) \leq \exp{\left(\frac{-\eps^2 \mu_l}{3}\right)}$.
\end{itemize}

\end{lem}

\begin{lem}[Hoeffding's Inequality] \label{lem:hoeffdingineq}

Let $X_1,\ldots,X_n$ be independent random variables such that $a_i \leq X_i \leq b_i$ and $X=\sum\limits_{i=1}^n X_i$. Then, for all $\delta >0$, 
$$ \pr\left(\size{X-\E[X]} \geq \delta\right) \leq  2\exp\left(\frac{-2\delta^2} {\sum\limits_{i=1}^{n}(b_i-a_i)^2}\right).$$

\end{lem}

\end{document}